\documentclass[a4paper,11pt]{article}
\pdfoutput=1
\usepackage{jheppub}
\usepackage{multirow}
\newcommand{\vev}[1]{\langle #1 \rangle}
\newcommand{\lessim}{\hspace{0.3em}\raisebox{0.4ex}{$<$}\hspace{-0.75em}\raisebox{-.7ex}{$\sim$}\hspace{0.3em}}
\renewcommand{\Re}{\text{Re\,}}
\renewcommand{\Im}{\text{Im\,}}
\newcommand{\Br}{\text{Br\,}}
\newcommand{\TeV}{\text{TeV}}
\newcommand{\GeV}{\text{GeV}}

\newcommand{\crit}{{\rm crit}}

\newcommand{\gtwo}{I\kern-.1em I\,}
\newcommand{\be}{\begin{equation}}
\newcommand{\ee}{\end{equation}}
\newcommand{\beq}{\begin{eqnarray}}
\newcommand{\eeq}{\end{eqnarray}}
\newcommand{\bpm}{\begin{pmatrix}}
\newcommand{\epm}{\end{pmatrix}}
\newcommand{\cl}{\, \rm C.L.}

\newcommand{\blackdiamond}{\raisebox{0.22ex}{\rotatebox[origin=c]{45}{\scriptsize $\blacksquare$}}\,}
\newcommand{\blackcircles}{\raisebox{-0.33ex}{\rotatebox[origin=c]{0}{\Large $\bullet$}}}

\title{126 GeV Higgs boson in the top-seesaw model}
\author[a]{Hidenori S. Fukano}
\author[b,c,d]{Kimmo Tuominen}
\affiliation[a]{Kobayashi-Maskawa Institute for the Origin of Particles and the Universe (KMI),
\\
Nagoya Universtiy, Furo-cho, Chikusa-ku, Nagoya 464-8602, Japan}
\affiliation[b]{Department of Physics, University of Jyv\"askyl\"a, \\
P.O.Box 35, FIN-40014 Jyv\"askyl\"a, Finland }
\affiliation[c]{
Helsinki Institute of Physics, \\
P.O.Box 64, FIN-00014 University of Helsinki, Finland
}
\affiliation[d]{Department of Physics and Astronomy, University of Southampton, Southampton, SO17 1BJ, UK }

\emailAdd{fukano@kmi.nagoya-u.ac.jp}
\emailAdd{kimmo.i.tuominen@jyu.fi}

\abstract{We consider a model of dynamical electroweak symmetry breaking built on the idea of top-seesaw mechanism. The model features a fourth generation of vector-like QCD quarks responsible for the origin of the top-seesaw mechanism and leading to the natural explanation of the large splitting between the top and bottom quark masses. Motivated by the LHC data on the couplings of the Higgs boson, we include the entire third generation of Standard Model matter fields into the model. We determine the low energy effective theory and the resulting low energy spectrum of states, and constrain the model parameters with constraints from the precision electroweak data and from the requirement of a light scalar state with quantum numbers of the Standard Model Higgs boson. Finally, we perform a global fit of the model parameters to the LHC Higgs data and show that the model is equally viable as the Standard Model itself, and predicts new states accessible at the LHC.}
\keywords{Electroweak symmetry breaking, top-seesaw, Higgs physics}
\begin{document}

\maketitle

\section{Introduction}
\label{intro}

The ATLAS \cite{Aad:2012tfa} and CMS collaborations \cite{Chatrchyan:2012ufa} have reported the discovery of a new boson with properties compatible with the Higgs boson in the Standard Model (SM) of the elementary particle physics. The implications of LHC data have been analyzed by several authors in literature, e.g. \cite{Azatov:2012bz,Carmi:2012yp,Espinosa:2012ir,Giardino:2012ww,Carena:2012xa,Corbett:2012dm,Giardino:2012dp,Ellis:2012hz,Carmi:2012in,Banerjee:2012xc,Joglekar:2012vc,ArkaniHamed:2012kq,Bonnet:2012nm,Wang:2012ts,Corbett:2012ja,Han:2013ic,Alanne:2013dra}. All these analyses underline the fact that SM provides an adequate description 
of all data within one standard deviation.

Despite this success, SM is necessarily an incomplete theory. It does not provide for a dark matter candidate or a mechanism for generation of matter-antimatter-asymmetry. Within the SM itself,  the hierarchical patterns of the observed masses of the matter fields remain completely unexplained. Analyzing the running of the SM coupling constants reveals that the SM with 126 GeV Higgs boson can persist up to extremely high energies \cite{Degrassi:2012ry}, and this motivates a discussion on the implications of naturality \cite{Farina:2013mla,Heikinheimo:2013fta,Craig:2013xia}. To interpret the situation there are, in broad terms, two possible 
alternatives: First, one may treat the observation of a light scalar, absence of any other new states and running of the Higgs quartic coupling towards zero as an indication that there really is only SM below scales $\Lambda\sim 10^{10}$ GeV, and the smallness of the Higgs mass is explained as a boundary condition of the matching of the SM onto a more complete theory which is scale invariant \cite{Heikinheimo:2013fta}.
Second, one takes the observation of a light scalar and no other new states as a strong constraint on the 
models built on the traditional naturality paradigm \cite{'tHooft:1979bh}. In this paper we will consider this latter viewpoint.

There are many models adapting naturalness as a guide beyond the SM in the market. Among them, several scenarios based on the strong coupling dynamics remain viable. A representative scenario is Technicolor (TC) \cite{Weinberg:1975gm,Susskind:1978ms} and its most promising realization, the walking TC scenario \cite{Holdom:1984sk,Yamawaki:1985zg,Akiba:1985rr,Appelquist:1986an,Sannino:2004qp,Dietrich:2005jn}, which is described by the gauge theory based on the near conformal dynamics. The new boson based on the walking TC scenario has been addressed by several authors \cite{Matsuzaki:2012gd,Elander:2012fk,Eichten:2012qb,Foadi:2012bb}. 

To explain the observed mass patterns of the known matter fields within the TC framework, a well-known approach is the extended TC (ETC) model building \cite{Dimopoulos:1979es,Eichten:1979ah}, in which the technicolored femions (technifermions) and the SM fermions are embedded into a larger gauge group (ETC gauge group). In ETC models, after the ETC gauge group breaks down to the TC gauge group and the technifermion condensation is triggered by the walking TC gauge dynamics, the SM fermions obtain their mass  from the technifermion condensates  via the four fermion interactions generated from the exchange of the massive ETC gauge bosons. If the ETC gauge group breaks sequentially, such ETC model may explain the observed mass hierarchies of the SM fermions \cite{Appelquist:1993sg,Appelquist:2003hn}. However, it might be hard to explain the large top quark mass, or more precisely, the large top-bottom mass splitting with keeping the consistency with the electroweak precision tests \cite{Peskin:1990zt,Peskin:1991sw}. To address this particular issue, an alternative to ETC, the top quark condensation model \cite{Miransky:1988xi,Miransky:1989ds,Nambu:1989jt,Marciano:1989xd,Marciano:1989mj,Bardeen:1989ds}, was proposed  in a form of the low energy effective model based on the gauged Nambu-Jona-Lasinio (NJL) model having large mass anomalous dimension $\gamma_m \simeq 2$ \cite{Miransky:1988xi,Miransky:1989ds}. The top quark condensation model generically predicts the existence of a SM Higgs-like bound state (the top-Higgs boson), which is a top quark composite, and whose mass ($m_h$) is related to the dynamical top quark mass ($m_{\rm dyn} = m_t \simeq 174 \,\GeV$) as $m^2_h = 4 m^2_{\rm dyn} = 4 m^2_t$. This top-Higgs boson is obviously not suitable to identify with the new boson with $m^2_h\simeq 126 \,\GeV$. However, we show that the top-Higgs boson with $\simeq 126\, \GeV$ can be realized 
in models based on top quark condensation; the particular model setup we have in mind is the top-seesaw model  \cite{Dobrescu:1997nm,Chivukula:1998wd,He:2001fz}; 
see also e.g. \cite{He:1999vp,Balazs:2012iz,Wang:2013jwa} for alternative models of this type.

In \cite{Fukano:2012qx,Fukano:2012nx} we have considered a model including walking TC and a top-seesaw model.  We pointed out that the top-Higgs boson can have $\simeq 126 \,\GeV$ by sharing the dynamical top quark mass with the TC sector, i.e. $m_{\rm dyn} < m_t$, while retaining the consistency with the constraints on precision electroweak observables. 
However, this possibility becomes highly constrained by the current LHC data, most notably by the 
results on the $\gamma\gamma$ decay channel of the Higgs boson  \cite{Aad:2012tfa,Chatrchyan:2012ufa}  and the vector boson fusion (VBF) production process of the Higgs boson \cite{ATLAS-CONF-2013-034,CMS-PAS-HIG-13-005}.

However there is another way to realize the top-Higgs with $\simeq 126 \,\GeV$. As the authors in \cite{Chivukula:1998wd} already pointed out, it might be possible to realize the top-Higgs boson with ${\cal O} (100 \,\GeV)$ in the top-seesaw model without sharing the top quark mass as in \cite{Fukano:2012nx}, by taking into account the condensation of the vector-like top-partner quark. In this paper our goal is to 
consider the top-seesaw model in detail from the viewpoint of the current LHC data to obtain clues for further model building in this framework or to see if this possibility is entirely ruled out. We find that the model is equally viable in light of the current data as the SM itself. Of course, this approach requires 
one to accept a certain level of fine tuning in order to accommodate a light Higgs particle into the theory. 
We will take this experimental result as basic input for our model, and investigate the consequences.

This paper organized as follows: In section \ref{TSS-review} we outline the top-seesaw model, and consider the third family fermions in the top-seesaw model proposed in \cite{Chivukula:1998wd}. In section \ref{EWPTandZbb} we discuss the constraint on the top-seesaw model from the current  electroweak precision test data including $R_b = \Gamma(Z\to b\bar{b})/\Gamma(Z \to \text{hadrons})$. In section \ref{VSLHC}, we discuss the top-Higgs in the top-seesaw model
in light of the current LHC data. Section \ref{summary} summarizes our results.

\section{Top-seesaw model}
\label{TSS-review}

\subsection{The effective Lagrangian}

In this paper our aim is not to construct a full ultraviolet complete model, but instead work directly with a low energy realization which allows to use current data to provide model building constraints in a bottom-up framework. As a starting point, we take effective four-fermion interactions, i.e. a Nambu-Jona-Lasinio (NJL) type model \cite{Miransky:1988xi,Miransky:1989ds,Nambu:1961tp,Chivukula:1998wd}, which we assume to be sourced by an underlying gauge theory with matter fields. This approach is in spirit of the topcolor model in \cite{Hill:1991at} or topcolor with $U(1)$ tilting mechanism \cite{Chivukula:1998wd}. 
Motivated by phenomenology, we start with the NJL Lagrangian describing the new physics and its effects on the full third generation of SM matter which is  defined at the cut-off scale $\Lambda$ as 
\beq
{\cal L}^{\rm TSS}_{\Lambda}
&=&
- \left[ \mu_{\chi \chi} \, \bar{\chi}_R \chi_L + \mu_{\chi t} \, \bar{t}_R \chi_L + \text{h.c.} \right]
\nonumber\\
&&
+
G_t \left( \bar{q}^\alpha_L \, t_R \right)\left( \bar{t}_R \, q^\alpha_L\right)
+
G_{q b} \left( \bar{q}^\alpha_L \, b_R \right)\left( \bar{b}_R \, q^\alpha_L\right)
+
G_{q \chi} \left( \bar{q}^\alpha_L \, \chi_R \right)\left( \bar{\chi}_R \, q^\alpha_L\right)
\nonumber \\
&&
+
G_{\chi \chi} \left( \bar{\chi}_L \, \chi_R \right)\left( \bar{\chi}_R \, \chi_L\right)
+
G_{\chi t} \left( \bar{\chi}_L \, t_R \right)\left( \bar{t}_R \, \chi_L\right)
+
G_{\chi b} \left( \bar{\chi}_L \, b_R \right)\left( \bar{b}_R \, \chi_L\right)
\nonumber\\
&&
+
G_2  \left[ 
\Bigl( \bar{q}^\alpha_L \, \chi_R \Bigr) (i\tau_2)^{\alpha \beta}\left( \bar{b}_R \, q^\beta_L \right)^c 
-
\Bigl( \bar{q}^\alpha_L \, b_R \Bigr)^c (i\tau_2)^{\alpha \beta}\left( \bar{\chi}_R \, q^\beta_L \right) 
\right]
\nonumber\\
&&
+
G_\tau  \left[ 
\Bigl( \bar{q}^\alpha_L \, \chi_R \Bigr) (i\tau_2)^{\alpha \beta}\left( \bar{\tau}_R \, l^\beta_L \right)^c 
-
\Bigl( \bar{l}^\alpha_L \, \tau_R \Bigr)^c (i\tau_2)^{\alpha \beta}\left( \bar{\chi}_R \, q^\beta_L \right) 
\right]
\,.\label{starting-NJLV1}
\eeq
In this equation $\chi_{L,R}$ is a vector-like QCD quark which transforms as singlet under the electroweak $SU(2)_L$ gauge symmetry and the charge of $\chi$ is the same as the top quark, $Q_\chi = +2/3$. 
The fields $q_L \equiv (t_L\, b_L)^T$ and $l_L \equiv (\nu_{\tau L}\, \tau_L)^T$ are the usual $SU(2)_L$ quark and lepton doublets and $\alpha,\beta = 1,2$ denote the $SU(2)$ indices, i.e. $q(l)^{1,2}_L$ corresponds to $t(\nu_\tau)_L,b(\tau)_L$, respectively. The second Pauli matrix is denoted by $\tau_2$, and the superscript $c$ stands for charge conjugation. The $SU(N_c)$ ($N_c=3$) color indices are contracted within each parenthesis. The four fermions couplings $G_A  (A = t, q b, q \chi, \chi t,\chi b,\chi\chi,2,\tau)$ are proportional to $1/\Lambda^2$ and arise from the physics above the cut-off scale $\Lambda$. 

The first line in Eq. (\ref{starting-NJLV1}) contains the $SU(2)_L$ singlet mass terms and $\mu_{\chi \chi}, \mu_{\chi t} >0$. The first, second and third lines in Eq. (\ref{starting-NJLV1}) are the same as the initial Lagrangian in \cite{Chivukula:1998wd}. In addition to them we consider the $G_2$- and $G_\tau$-terms (fourth and fifth lines in Eq. (\ref{starting-NJLV1})), since we want to describe also the bottom quark and tau lepton masses via the top-seesaw model. Similar interactions were considered in e.g. \cite{Miransky:1988xi,Miransky:1989ds}. %
In the top-seesaw model the condensate to trigger the electroweak symmetry breaking (EWSB) is $\vev{\bar{t}_R \chi_L} \neq 0$ which is different from the original top quark condensation model where $\vev{\bar{t}_R t_L} \neq 0$ triggers the EWSB. We do not allow for the tau condensation, i.e. we impose $\vev{\bar{\tau}_R \tau_L} = 0$ 
and therefore we do not include the four fermion interactions of the form $(\bar{l}_L \tau_R)^2$ explicitly in Eq. (\ref{starting-NJLV1}). The Lagrangian in Eq. (\ref{starting-NJLV1}) is of course invariant under the SM gauge symmetry.

To describe the physics at $\mu (< \Lambda)$, we use the large-$N_c$ fermion loop approximation \cite{Bardeen:1989ds}. We discuss only the essential elements of the analysis and show the final result here. For the details of the derivation, see Appendix \ref{efflagrderivation}. 

It is convenient first to diagonalize $G_{q \chi}$, $G_{q b}$ and $G_2$-terms in 
Eq.(\ref{starting-NJLV1}) \cite{Harada:1990wg}. The resulting eigenvalues 
$G_{\chi,b}\,,(G_\chi > G_b)$ are
\beq
G_{\chi , b} = \frac{1}{2} \left[ G_{q \chi} + G_{q b} \pm  \sqrt{(G_{q \chi} - G_{q b})^2 + 4G^2_2}\right]
\,,\label{def-ev-4f}
\eeq
and the mixing angle is determined by by $(0 \leq \theta \leq \pi/2)$
\beq
\cos^2\theta = \frac{1}{2} \left[ 1 + \frac{G_{q \chi} - G_{q b}}{\sqrt{(G_\chi - G_b)^2 + 4G^2_2}}\right]
\,,\,
\sin^2\theta = \frac{1}{2} \left[ 1 - \frac{G_{q \chi} - G_{q b}}{\sqrt{(G_\chi - G_b)^2 + 4G^2_2}}\right]
\,.\label{diag-matrix}
\eeq
To achieve desired symmetry breaking patterns and quark mass phenomenology, several conditions on the
four fermion couplings are imposed. First, to relate the bottom quark mass and the condensate 
$\vev{\bar{\chi}_R q_L} \neq 0$, we assume $G_b = 0$, i.e. we impose $G_{q \chi} G_{q b} - G^2_2 =0$ in Eq. (\ref{starting-NJLV1}).  
We also impose on $G_t$, $G_{\chi}$ and $G_{\chi b}$ the following criticality conditions,
\beq
0 < G_t < G_\crit < G_{\chi}
\quad ,\quad
0 < G_{\chi b} < G_\crit 
\quad , \quad
0 < G_{\tau} \ll G_\crit
\,,\label{criticality-GABV1}
\eeq
where $G^{-1}_\crit \equiv N_c \Lambda^2 /(8 \pi^2)$ is the critical four fermion coupling. The first condition in Eq.(\ref{criticality-GABV1}) means that the dominant contribution to the EWSB arises from the condensate 
$\vev{\bar{\chi}_R \, q_L} \neq 0$. The second condition in Eq.(\ref{criticality-GABV1}) is required to preserve the $U(1)_{\rm e.m.}$ gauge symmetry on the vacuum. The third condition in Eq.(\ref{criticality-GABV1}), is required to forbid the tau lepton condensation, and we will treat $G_\tau$ as a parameter to relate the tau lepton mass and the condensate $\vev{\bar{\chi}_R \, q_L} \neq 0$. For $G_{\chi \chi},G_{\chi t} (>0)$, we do not impose any criticality conditions and we treat them as free parameters.

The low energy effective theory is defined in terms of composite fields corresponding to the fermion bilinears appearing in Eq. (\ref{starting-NJLV1}) in the relevant channels as discussed above. The electroweak doublets are $\Phi_t\sim (\bar{q}_L t_R)$ and $\Phi_\chi\sim (\bar{q}_L\chi_R)$, and here are furthermore three electroweak singlet fields $\phi_{\chi f}\sim \bar{\chi}_L f_R$, where $f=\chi$, $t$ or $b$.

Applying the large-$N_c$ fermion loop approximation, we obtain the effective Lagrangian valid for $\mu < \Lambda$,
\beq
{\cal L}_{\mu < \Lambda}
&=&
\left[\left| D_\mu \Phi_t \right|^2 + \left| D_\mu \Phi_\chi \right|^2 + \sum_{f=t,b,\chi} \left| D_\mu \phi_{\chi f} \right|^2 \right]
\nonumber\\
&&
+y \left[
\bar{\psi}^\alpha_L \Phi^\alpha_t t_R
+
\cos \theta \, \bar{\psi}^\alpha_L \Phi^\alpha_\chi \chi_R
+
\sin \theta \, \bar{\psi}^\alpha_L \tilde{\Phi}^\alpha_\chi b_R
+
r_\tau \bar{l}^\alpha_L \tilde{\Phi}^\alpha_\chi \tau_R
+
\sum_{f=t,b,\chi} \bar{\chi}_L \phi_{\chi f} f_R
+
\text{h.c.}
\right]
\nonumber\\
&&
- V(\Phi,\phi)
\,,\label{eff-LagV1}
\eeq
where $y$ is given by
\beq
y 
=\frac{4\pi}{\sqrt{N_c \ln(\Lambda^2/\mu^2)}}
\,,\label{TSS-yukawa}
\eeq
and $r_\tau \equiv G_\tau/G_\chi$ and $\tilde{\Phi}^\alpha_\chi \equiv (-i\tau_2)^{\alpha \beta}\Phi^{*\beta}_\chi$.  The potential $V(\Phi,\phi)$ is 
\beq
V(\Phi,\phi)
&=&
\left[ 
M^2_t \left| \Phi_t\right|^2 
+
M^2_\chi \left| \Phi_\chi\right|^2 
+
\sum_{f=t,b,\chi} M^2_{\chi f} \left| \phi_{\chi f}\right|^2
\right]
\nonumber\\
&&
+ C_{\chi t} \left[ \phi_{\chi t} + \phi^\dagger_{\chi t} \right]
+ C_{\chi \chi} \left[ \phi_{\chi \chi} + \phi^\dagger_{\chi \chi} \right]
\nonumber\\
&&
+\frac{\lambda}{2}
\left[ 
\begin{aligned}
&
\left( \Phi^\dagger_t \Phi_t + \phi^\dagger_{\chi t} \phi_{\chi t}  \right)^2
\\
&
+
\left( \sin^2\theta \Phi^\dagger_\chi \Phi_\chi + \phi^\dagger_{\chi b} \phi_{\chi b}  \right)^2
+
\left( \cos^2\theta \Phi^\dagger_\chi \Phi_\chi + \phi^\dagger_{\chi \chi} \phi_{\chi \chi}  \right)^2
\\
&
+
2 \left| \sin \theta \Phi^\dagger_t \tilde{\Phi}_\chi  + \phi^\dagger_{\chi t} \phi_{\chi b}\right|^2
+
2 \left| \cos \theta \Phi^\dagger_\chi \Phi_t  + \phi^\dagger_{\chi \chi} \phi_{\chi t}\right|^2
\\
&
+
2 \left| \sin\theta \cos \theta \tilde{\Phi}^\dagger_\chi \Phi_\chi + \phi^\dagger_{\chi b} \phi_{\chi \chi}\right|^2
\end{aligned}
\right]
\,,\label{eff-potential-0V1}
\eeq
where the parameters $M^2_A $ ($A=t,\chi,\chi t, \chi b, \chi \chi$), $C_B$ ($B={\chi t, \chi\chi}$) and $\lambda$ are given by

\beq
&&
M^2_A 
= \frac{2\Lambda^2}{\ln(\Lambda^2/\mu^2)} \left( \frac{1}{g_A}-1\right)
\,,\label{M2-finV1}\\
&&
C_B 
= \mu_{f f'} \frac{\Lambda^2}{2\pi g_B} \sqrt{\frac{N_c}{\ln(\Lambda^2/\mu^2)}} 
\,,\label{tadpole-finV1}\\
&&
\lambda 
= \frac{32 \pi^2}{N_c \ln(\Lambda^2/\mu^2)} = 2y^2 
\,.\label{lambda-finV1}
\eeq
Here we have defined $g_A=\left[ N_c\Lambda^2/(8\pi^2) \right]G_A$.
\subsection{Mass spectrum of the fermions and composite scalars}

Next we consider the low energy mass spectrum of states arising from the Lagrangian 
Eq. (\ref{starting-NJLV1}). 
We consider the effective potential $V(\Phi,\phi)$ in Eq.(\ref{eff-potential-0V1}) together with Eqs. (\ref{M2-finV1}), (\ref{tadpole-finV1}), (\ref{lambda-finV1}). %
We parametrize the $SU(2)_L$ doublet scalar fields $\Phi$ as
\beq
\Phi_t = \bpm  \dfrac{1}{\sqrt{2}}\varphi^0_{tt} \\[2ex] \varphi^-_{bt} \epm
\,,\,
\Phi_\chi = \bpm \dfrac{1}{\sqrt{2}}\varphi^0_\chi \\[2ex] \varphi^-_\chi \epm
\,,\label{parametrize-Higgs}
\eeq
where each component $\varphi$ is a complex scalar field. For $SU(2)_L$ singlet scalars we write $\phi = (1/\sqrt{2}) \left(\Re\phi + i \Im \phi\right)$. Requiring the physical vacuum to preserve the $U(1)_{\rm e.m.}$ gauge symmetry and the CP-symmetry, the vacuum expectation values for the doublet fields $\varphi$
must be real and electrically neutral,
and the vacuum expectation value for the $SU(2)_L$ singlet scalar fields $\phi$ should also be real and 
furthermore satisfy $\vev{\phi_{\chi b}}=0$. Thus, out of the two complex $SU(2)_L$ doublet scalar fields and three complex $SU(2)_L$ singlet scalar fields, there are four possible non-zero vacuum expectation values:
\beq
\vev{\Re \varphi^0_{tt}} \equiv \tilde{v}_{tt}
\,,\,
\vev{\Re \varphi^0_\chi} \equiv \tilde{v}_\chi
\,,\,
\vev{\Re \phi_{\chi \chi}} \equiv \tilde{v}_{\chi \chi}
\,,\,
\vev{\Re \phi_{\chi t}} \equiv\tilde{v}_{\chi t}
\,,\label{vev}
\eeq
which are determined by the minimization condition for the effective potential $V(\Phi,\phi)$ in Eq. (\ref{eff-potential-0V1}). The minima are detemined from
\beq
V_0
&\equiv&
\left. V(\Phi,\phi) \right|_{\vev{\Phi},\vev{\phi}}
\nonumber\\
&=&
\sqrt{2} C_{\chi t} \tilde{v}_{\chi t} + \sqrt{2} C_{\chi \chi} \tilde{v}_{\chi \chi}
+ 
\frac{1}{2}M^2_\chi \tilde{v}^2_\chi + \frac{1}{2}M^2_t \tilde{v}^2_{tt}
+
\frac{1}{2}M^2_{\chi \chi} \tilde{v}^2_{\chi \chi} + \frac{1}{2}M^2_{\chi t} \tilde{v}^2_{\chi t} 
\nonumber\\
&&
+
\frac{\lambda}{8}
\left[
\left( \tilde{v}^2_{tt} + \tilde{v}^2_{\chi t} \right)^2
+ \tilde{v}^4_\chi \sin^4\theta
+ \left(\tilde{v}^2_\chi  \cos^2\theta + \tilde{v}^2_{\chi \chi} \right)^2
+ 2 \left(\tilde{v}_{tt} \tilde{v}_\chi \cos \theta + \tilde{v}_{\chi \chi} \tilde{v}_{\chi t} \right)^2
\right]
\,.\label{eff-potential-vev}
\eeq
Let us comment on the phase of $\tilde{v}$ here. 
First, the phases of $\tilde{v}_{\chi \chi},\tilde{v}_{\chi t}$ are forced to be negative by the tadpole terms since the coefficients $C_{\chi t, \chi\chi}$ are  positive (see Eq. (\ref{tadpole-finV1})). Moreover, to minimize the effective potential Eq. (\ref{eff-potential-0V1}) the phase of $\tilde{v}_{tt} \tilde{v}_\chi$ should be negative and we choose $\tilde{v}_{tt}>0$ and $\tilde{v}_{t\chi} <0$ here. Based on these facts, we rewrite the barred quantities in Eqs. (\ref{vev}) as 
\beq
\tilde{v}_{tt} = v_{tt}
\,,\,
\tilde{v}_\chi = -v_\chi
\,,\,
\tilde{v}_{\chi \chi} = -v_{\chi \chi}
\,,\,
\tilde{v}_{\chi t} = -v_{\chi t}
\,,
\label{vev-fin}
\eeq
where all $v$ are positive. We define the following ratios 
\beq
\frac{v_{tt}}{v_{\chi\chi}} = ab\epsilon
\,,\quad
\frac{v_\chi}{v_{\chi\chi}} = \epsilon
\,,\quad
\frac{v_{\chi t}}{v_{\chi\chi}} = b
\,,\label{ratio-vev}
\eeq
and we impose the seesaw condition \cite{Chivukula:1998wd}:
\beq
0 < \epsilon < b < 1
\,,\quad
0 < a \ll \frac{1}{\epsilon}
\,,\quad
\epsilon \ll 1
\,,\label{seesaw-condition}
\eeq
where $a$, $b$ and $\epsilon$ are dimensionless parameters. 

From the minimization conditions on $v_{tt},v_\chi$ 
together with Eq.(\ref{ratio-vev}), we obtain $M^2_{t,\chi}$ as a function of
$\lambda,v_{\chi \chi},a,b,\epsilon,\theta$ :
\beq
M^2_t &=& 
\frac{1}{2} \lambda v^2_{\chi \chi} \left[ 
\frac{1}{a}\cos \theta - b^2 - \epsilon^2 \cos^2\theta - a^2b^2\epsilon^2
\right]
\,,\label{mini-tt}\\[1ex]
-M^2_\chi &=& 
\frac{1}{2} \lambda v^2_{\chi\chi} \left[ 
\begin{aligned}
&
\cos^2\theta - ab^2 \cos\theta 
\\
&
+ \epsilon^2 (\cos^4\theta + \sin^4\theta) + a^2b^2\epsilon^2 \cos^2\theta
\end{aligned}
\right]
\,,\label{mini-chi}
\eeq
which should satisfy the criticalitity condition $M^2_t >0$ and $-M^2_\chi >0$ (see Eq. (\ref{criticality-eff-1}). Moreover, we obtain the lower bound for $M^2_{\chi t, \chi\chi}$ by considering the minimization conditions for $v_{\chi t}$ and $v_{\chi\chi}$ together with Eq. (\ref{tadpole-finV1}) :
\beq
M^2_{\chi t}  &>& 
-\frac{1}{2}\lambda v^2_{\chi \chi} \left[ 1 + b^2 + a^2 b^2 \epsilon^2 - a\epsilon^2 \cos\theta \right]
\nonumber\\
&=& 
M^2_\chi \frac{1+b^2}{\cos^2\theta - ab^2\cos\theta} \left[ 1+{\cal O}(\epsilon^2)\right]
\,,\label{mini-chit}\\[1ex]
M^2_{\chi \chi}  &>& 
-\frac{1}{2}\lambda v^2_{\chi \chi} \left[ 1 + b^2 + \epsilon^2 \cos^2\theta - ab\epsilon^2 \cos\theta \right]
\nonumber\\
&=&
M^2_\chi \frac{1+b^2}{\cos^2\theta - ab^2\cos\theta} \left[ 1+{\cal O}(\epsilon^2)\right]
\,.\label{mini-chichi}
\eeq
The above discussion parallels the discussion in \cite{Chivukula:1998wd}, except that we have included
the $G_2$- and $G_\tau$-terms in the Lagrangian in Eq.(\ref{starting-NJLV1}). 
If we set $\cos \theta =1$ and $G_\tau = 0$, one can see easily that Eqs.(\ref{mini-tt}), (\ref{mini-chi}), (\ref{mini-chit}) and (\ref{mini-chichi}) reduce to the results in \cite{Chivukula:1998wd}. 

After these preliminaries on the effective potential, we can derive the fermion mass spectrum for $t,b,\chi,\tau$. These fermion masses originate from the Yukawa terms in Eq. (\ref{eff-LagV1}),
\beq
{\cal L}^{\rm mass}_{\rm fermion}
=
-m_\tau \bar{\tau}_L \tau_R
-m_b \bar{b}_L b_R 
-
\bpm \bar{t}_L & \bar{\chi_L} \epm
{\cal M}_{t\chi}
\bpm t_R \\ \chi_R \epm
+
\text{h.c.}
\,,\label{mass-term}
\eeq
where $m_\tau\,$,$\,m_b\,$ and $\,{\cal M}_{t\chi}$ are given by
\beq
m_\tau = \frac{y}{\sqrt{2}} v_\chi r_\tau
\,,\,
m_b = \frac{y}{\sqrt{2}} v_\chi \sin \theta
\,,\,
{\cal M}_{t\chi}
\equiv
\bpm m_{tt} & m_{t\chi} \\ m_{\chi t} & m_{\chi \chi} \epm
=
\frac{y}{\sqrt{2}}\bpm -v_t & v_\chi \cos\theta\\ v_{\chi t} & v_{\chi \chi} \epm
\,.\label{mass-matrix-top}
\eeq
The $t$ and $\chi$ masses ($m_{t,\chi}$) are given by the eigenvalues of ${\cal M}^\dagger_{t\chi} {\cal M}_{t\chi}$ or ${\cal M}_{t\chi}{\cal M}^\dagger_{t\chi}$ in which smaller (larger) eigenvalue corresponds to $m^2_t$ $(m^2_\chi)$,
\beq
m^2_t
&=&
\frac{m^2_{t\chi}}{\cos^2\theta} \frac{b^2(1+a)^2}{1+b^2}\left[ 1 + {\cal O}(\epsilon^2)\right]
\,,\label{top-mass}\\
m^2_\chi
&=&
\frac{m^2_{t\chi}}{\cos^2\theta} \frac{1+b^2}{\epsilon^2}\left[ 1 + {\cal O}(\epsilon^2)\right]
\,,\label{chi-mass}
\eeq
where $m_{t\chi}= yv_\chi \cos\theta/\sqrt{2}$.
As already mentioned, the validity of the analysis requires $\Lambda/\mu = \Lambda/m_\chi > m_\chi/m_t$. From the above equations we deduce that this validity condition is equivalent to  
\beq
\epsilon > \frac{1+b^2}{b(1+a)} \left( \frac{\Lambda}{m_\chi}\right)^{-1}
\,.\label{epsilon-lowerbound}
\eeq
In the present top-seesaw model, the bottom quark mass $m_b$ and tau lepton mass $m_\tau$ as a function of the dynamical mass $m_{t\chi}$ are given by
\beq
m_\tau = m_{t \chi} r_\tau
\,,\,
m_b = m_{t\chi} \tan\theta
\,.\label{bottom-mass}
\eeq
On the other hand, the vacuum expectation value for the EWSB ($v_{\rm EW} = 246 \,\GeV$) is given by
\beq
 v^2_{\rm EW} \equiv v^2_{tt}+ v^2_\chi
 \,,\label{def-EWvev}
\eeq
which gives a constraint on the dynamical mass $m^2_{t\chi}$ as
\beq
m^2_{t\chi}
&=&
\frac{8\pi^2 v^2_{\rm EW}}{N_c} \cos^2\theta \left[\left(1+a^2b^2\right)\ln\frac{\Lambda^2}{\mu^2}\right]^{-1} 
\nonumber\\
&\simeq&
(1.26\,\TeV)^2 
\times
\cos^2\theta
\times
\left( \frac{v_{\rm EW}}{246 \,\GeV}\right)^2 
\times
\left[\left(1+a^2b^2\right)\ln\frac{\Lambda^2}{\mu^2}\right]^{-1} 
\,.\label{dynamical-mass}
\eeq
Thus, to obtain realistic top quark mass we find that $a,b$ should satisfy
\beq
\frac{b^2(1+a)^2}{(1+b^2)^2(1 + a^2b^2)} 
&=&
\frac{N_c m^2_t}{8\pi^2 v^2_{\rm EW}} \ln \frac{\Lambda^2}{\mu^2}
\nonumber\\
&\simeq&
(0.14)^2 
\times 
\left( \frac{m_t}{175(\GeV)}\right)^2 \left( \frac{246\,\GeV}{v_{\rm EW}}\right)^2
\times
\ln\frac{\Lambda^2}{\mu^2}  
\,.\label{top-constraint}
\eeq
To obtain the realistic bottom quark mass, $\sin \theta$ should satisfy
\beq
\sin^2\theta
&=&
\frac{N_c m^2_b}{8\pi^2 v^2_{\rm EW}} 
\left[\left(1+a^2b^2\right)\ln\frac{\Lambda^2}{\mu^2}\right]
\nonumber\\
&\simeq&
(0.0032)^2 
\times 
\left( \frac{m_b}{4(\GeV)}\right)^2 
\times
\left( \frac{246\,\GeV}{v_{\rm EW}}\right)^2
\times
\left[\left(1+a^2b^2\right)\ln\frac{\Lambda^2}{\mu^2}\right]
\,,
\eeq
and, finally, to obtain the realistic tau lepton mass, $G_\tau/G_\chi$ should satisfy
\beq
r^2_\tau
&=&
\frac{N_c m^2_\tau}{8\pi^2 v^2_{\rm EW}} 
\left[\left(1+a^2b^2\right)\ln\frac{\Lambda^2}{\mu^2}\right]
\nonumber\\
&\simeq&
(0.0013)^2 
\times 
\left( \frac{m_\tau}{1.7(\GeV)}\right)^2 
\times
\left( \frac{246\,\GeV}{v_{\rm EW}}\right)^2
\times
\left[\left(1+a^2b^2\right)\ln\frac{\Lambda^2}{\mu^2}\right]
\,.
\eeq
If $\Lambda/\mu$ is not very large, e.g. $\Lambda/\mu \sim {\cal O}(10-100)$, and $\ln (\Lambda^2/\mu^2) \simeq {\cal O}(2-3)$, and we set $a \simeq {\cal O}(1)$, then the typical values of $b$, $\sin \theta$ and $r_\tau$ are 
\beq
b \simeq 0.15 - 0.22
\,,\,
\sin \theta \simeq 0.007-0.01
\,,\,
r_\tau \simeq 0.003-0.004.
\label{typical-values}
\eeq
This implies that $b^2 \lessim 0.05$ and $\sin^2\theta \lessim 0.0001$, so it is reasonable to neglect higher order contributions in $b$ and $\sin\theta$. Given the above estimates for the parameters, the validity of our analysis requires that $\epsilon$ should satisfy $\epsilon \gtrsim 0.34 $ for $\Lambda/\mu = 10$ and $\epsilon \gtrsim 0.024$ for $\Lambda/\mu = 100$ from Eq. (\ref{epsilon-lowerbound}). 

For the later purposes, it is useful to rewrite the the top ($t$) and its vector-like partner ($\chi$) in their mass basis instead of the gauge basis:
\beq
&&
\bpm t^{(g)}_L \\ \chi^{(g)}_L \epm
=
\bpm c^t_L & s^t_L \\ -s^t_L & c^t_L\epm
\bpm t^{(m)}_L \\ \chi^{(m)}_L \epm
\equiv
U_L \bpm t^{(m)}_L \\ \chi^{(m)}_L \epm
\,,\label{rotate-L}
\\[1ex]
&&
\bpm t^{(g)}_R \\ \chi^{(g)}_R \epm
=
\bpm -c^t_R & s^t_R \\ s^t_R & c^t_R\epm
\bpm t^{(m)}_R \\ \chi^{(m)}_R \epm
\equiv
U_R \bpm t^{(m)}_R \\ \chi^{(m)}_R \epm
\,,\label{rotate-R}
\eeq
where $c^t_L \equiv \cos \theta^t_L\,,\cdots$ and the superscripts $(g)$ and $(m)$ imply the gauge basis and the mass basis, respectively. Hereafter we will drop these superscripts to simplify notation. The mixing angles $c^t_L$ and $c^t_R$ are given by
\beq
&&
c^t_L 
= 
\frac{1}{\sqrt{2}} \left[ 
1 + \frac{m^2_{\chi \chi} + m^2_{\chi t} - m^2_{tt} -m^2_{t\chi}}{m^2_\chi - m^2_t}
\right]^{1/2}
\,,\label{def-cLt}\\
&&
c^t_R
= 
\frac{1}{\sqrt{2}} \left[ 
1 + \frac{m^2_{\chi \chi} + m^2_{t \chi} - m^2_{tt} -m^2_{\chi t}}{m^2_\chi - m^2_t}
\right]^{1/2}
\,,\label{def-cRt}
\eeq
where $m_{tt,\chi\chi, \chi t, t\chi}$ and $m_{t,\chi}$ are given in Eqs. (\ref{mass-matrix-top}), (\ref{top-mass}) and (\ref{chi-mass}). %

Then we discuss the composite scalar boson mass spectrum in this model. %
There are three charged composite scalar fields $\varphi^\pm_{bt},\varphi^\pm_\chi$ and $\phi^\pm_{b\chi}$. The first two of these scalars originate from the $SU(2)_L$ doublets in Eq. (\ref{parametrize-Higgs}) and the third scalar from the $SU(2)_L$ singlet. To the lowest order in $\epsilon$, $b$ and $\sin\theta$, their mass eigenstates are given by
\beq
\bpm G^\pm \\ H^\pm \\ H^\pm_{\chi b} \epm
=
\bpm
\cos \alpha & -\sin \alpha & 0
\\
\sin\alpha &\cos \alpha & 0
\\
0 & 0 & 1
\epm
\bpm \varphi^\pm_\chi \\ \varphi^\pm_{bt} \\ \phi^\pm_{b\chi} \epm
\,,
\eeq
where 
\beq
\cos \alpha \equiv \frac{1}{\sqrt{1+a^2b^2}}
\,,\,
\left( 0 \leq \alpha \leq \frac{\pi}{2} \right)
\,.
\eeq
The state $G^\pm$ is the massless would-be Nambu-Goldstone boson which is absorbed in the $W^\pm$. The nonzero mass eigenvalues are given by 
\beq
&&
M^2_{H^\pm} =
\frac{2m^2_{t\chi}}{\cos\theta} \frac{1+a^2b^2}{a\epsilon^2}
\simeq {\cal O}((10\dots 100\, \TeV)^2)
\,,\label{charged-Higgs-mass}
\\ 
&&
M^2_{H^\pm_{\chi b}}=
M^2_{\chi b} + \frac{2m^2_{t\chi}}{\cos^2\theta} \frac{1+b^2}{\epsilon^2}
\simeq \Lambda^2
\,.\label{singlet-charged-Higgs-mass}
\eeq
Here we have used Eqs. (\ref{dynamical-mass}) and (\ref{typical-values}) and $\epsilon\sim {\cal O}(0.1\dots 0.01)$ to obtain the last approximation in Eq.(\ref{charged-Higgs-mass}). As the Eqs. (\ref{charged-Higgs-mass}) and (\ref{singlet-charged-Higgs-mass}) show,  the charged composite scalars are very heavy 
and decouple from the low energy phenomenology at sub-TeV scales. 

In addition to the charged states there are four CP-even neutral composite scalar fields, $\Re\varphi_{tt}$, $\Re\varphi_\chi$, $\Re \phi_{\chi t}$ and $\Re \phi_{\chi\chi}$. First two of these scalars originate from the $SU(2)_L$ doublets in Eq. (\ref{parametrize-Higgs}), and the last two scalars are originated form the $SU(2)_L$ singlet. To the lowest order in $\epsilon$, $b$ and $\sin\theta$ 
their mass eigenstates are given by\footnote{Note that all states are rotated by the same angle, $\alpha$, even if the scalar sector is similar to a two-Higgs doublet model. This holds only to the lowest order of 
$\epsilon$.}
\beq
\bpm h^0 \\ H^0 \\ H^0_{\chi t} \\ H^0_{\chi \chi}\epm
=
\bpm
\cos \alpha & -\sin \alpha & 0 & 0
\\
\sin\alpha &\cos \alpha & 0 & 0
\\
0 & 0 & 1 & 0
\\
0 & 0 & 0 & 1
\epm
\bpm \Re\varphi_\chi \\ \Re\varphi_{tt} \\ \Re\phi_{\chi t} \\ \Re\phi_{\chi\chi} \epm
\,,\label{mixing-CPevenHiggs}
\eeq
with the corresponding mass eigenvalues given by
\beq
m^2_{h^0} 
&=&
4m^2_{t\chi} \cos^2\theta  \frac{M^2_{\chi\chi}-M^2_\chi}{M^2_{\chi\chi}-3M^2_\chi} 
\,,\label{mh0}
\\
m^2_{H^0}
&=&
\frac{2m^2_{t\chi}}{\cos \theta} \frac{1+a^2b^2}{a\epsilon^2}
\,,\label{mH0}\\
m^2_{H^0_{\chi t}}
&=& 
\frac{2m^2_{t\chi}}{\epsilon^2} \left[ \frac{1}{\cos^2\theta} + \frac{M^2_{\chi t}}{-M^2_\chi}\right]
\,,\label{mHchit0}\\
m^2_{H^0_{\chi \chi}}
&=& 
\frac{2m^2_{t\chi}}{\epsilon^2} \left[ \frac{3}{\cos^2\theta} + \frac{M^2_{\chi \chi}}{-M^2_\chi}\right]
\,.\label{mHchichi0}
\eeq
From these we see that $m^2_{H^0} = m^2_{H^\pm} \simeq {\cal O}((10\dots 100\, \TeV)^2)$, as estimated above, so $H^0$ decouples from low energy physics. The other neutral CP-even composite scalars can be light. %
Finally there are four CP-odd neutral composite scalar particles 
$\Im\varphi_{tt}, \Im\varphi_\chi, \Im \phi_{\chi t}$ and $\Im \phi_{\chi\chi}$. First two of these originate 
from the $SU(2)_L$ doublets in Eq.(\ref{parametrize-Higgs}) and the last two originate form the 
$SU(2)_L$ singlet. To the lowest order in $\epsilon$, $b$ and $\sin\theta$ their mass eigenstates are given by
\beq
\bpm G^0 \\ A^0 \\ A^0_{\chi t} \\ A^0_{\chi \chi}\epm
=
\bpm
\cos \alpha & -\sin \alpha & 0 & 0
\\
\sin\alpha &\cos \alpha & 0 & 0
\\
0 & 0 & 1 & 0
\\
0 & 0 & 0 & 1
\epm
\bpm \Im\varphi_\chi \\ \Im\varphi_{tt} \\ \Im\phi_{\chi t} \\ \Im\phi_{\chi\chi} \epm
\,.\label{mixing-CPoddHiggs}
\eeq
The state $G^0$ is the massless would-be Nambu-Goldstone boson which is absorbed in the $Z^0$. The three nonzero mass eigenvalues are given by 
\beq
m^2_{A^0}
&=&
\frac{2m^2_{t\chi}}{\cos \theta} \frac{1+a^2b^2}{a\epsilon^2}
\,,\label{mA0}\\
m^2_{A^0_{\chi t}}
&=& 
\frac{2m^2_{t\chi}}{\epsilon^2} \left[ \frac{1+b^2}{\cos^2\theta} + \frac{M^2_{\chi t}}{-M^2_\chi}\left( 1-\frac{ab^2}{\cos \theta}\right)\right]
\,,\label{mAchit0}\\
m^2_{A^0_{\chi \chi}}
&=& 
\frac{2m^2_{t\chi}}{\epsilon^2} \left[ \frac{1+b^2}{\cos^2\theta} + \frac{M^2_{\chi \chi}}{-M^2_\chi}\left( 1-\frac{ab^2}{\cos \theta}\right)\right]
\,.\label{mAchichi0}
\eeq
From these we again see that $m^2_{A^0} = m^2_{H^\pm} \simeq {\cal O}((10\dots 100\, \TeV)^2)$, so $A^0$ decouples from the low energy physics while the other neutral CP-odd composite scalars  can be light. %

Hence, the  low energy spectrum contains five potentially light composite scalars. Three of these are 
CP-even neutral scalars ($h^0,H^0_{\chi t},H^0_{\chi \chi}$) and two are CP-odd neutral scalars ($A^0_{\chi t},A^0_{\chi \chi}$). Again, in the limit $\cos \theta =1$, the above results reduce to the results of \cite{Chivukula:1998wd}. %
Concentrating on the five potentially light composite Higgs bosons, we furthermore note the following:  
From Eqs.(\ref{mHchichi0}), we find that $H^0_{\chi \chi}$ can not be $\sim {\cal O}(100 \,\GeV)$. For the four neutral scalars ($h^0,H^0_{\chi t}, A^0_{\chi t}, A^0_{\chi\chi}$), we rewrite Eqs. (\ref{mh0}), (\ref{mHchit0}), (\ref{mAchit0}) and (\ref{mAchichi0}) by using $g$ defined below Eq. (\ref{lambda-finV1}) as
\beq
m^2_{h^0} 
&=&
4m^2_{t\chi} \frac{g_\chi - g_{\chi\chi}}{g_\chi - 3g_{\chi \chi}+2g_\chi g_{\chi\chi}} \cos^2\theta
\,,\label{mh0-g}
\\
m^2_{H^0_{\chi t}}
&=&
\frac{2m^2_{t\chi}}{\epsilon^2} \left[ \frac{1}{\cos^2\theta} - \frac{g_\chi(g_{\chi t}-1)}{g_{\chi t}(g_\chi-1)}\right]
\,,\label{mHchit0-g}
\\
m^2_{A^0_{\chi t}} 
&=& 
\frac{2m^2_{t\chi}}{\epsilon^2} (1+b^2 )
\left[ 
\frac{1}{\cos^2\theta}-\frac{ \cos^2\theta - ab^2\cos\theta}{(1+b^2)\cos^2\theta } \frac{g_\chi(g_{\chi t}-1)}{g_{\chi t}(g_\chi-1)}
\right]
\,,\label{mAchit0-g}
\\
m^2_{A^0_{\chi \chi}} 
&=& 
\frac{2m^2_{t\chi}}{\epsilon^2} (1+b^2 )
\left[ 
\frac{1}{\cos^2\theta}-\frac{ \cos^2\theta - ab^2\cos\theta}{(1+b^2)\cos^2\theta } \frac{g_\chi(g_{\chi \chi}-1)}{g_{\chi \chi}(g_\chi-1)}
\right]
\,.\label{mAchichi0-g}
\eeq
Hence, there are four Higgs bosons to identify the new bosons at $\simeq 126 \,\GeV$ \cite{Aad:2012tfa,Chatrchyan:2012ufa} in the present top-seesaw model, and restricting to CP-even states leaves only $H^0_{\chi t}$ and $h^0$ as viable candidates for the boson discovered at the LHC.

As we discussed in Sec. \ref{intro}, the requirement of light scalars may necessitate some fine tuning. 
In our model this appears for the values of couplings $g_A$ relative to the critical value $g_A=1$.
If $H^0_{\chi t}, A^0_{\chi t}, A^0_{\chi \chi}$ should have $\sim {\cal O}(100\,\GeV)$ mass,
we should require $g_{\chi t} \to 1$ and/or $g_{\chi \chi} \to 1$.  To estimate the level of fine tuning, 
we consider $\epsilon=0.1$ and using the estimates in Eqs. (\ref{dynamical-mass}) and (\ref{typical-values}) we find that $m_{h^0} \simeq 126 \,\GeV$ if we allow for fine tuning of the order of 
${\cal O}(126^2/(4m^2_{t\chi})) \simeq 10^{-3}$ for the couplings. Similarly, we find that $m_{H^0_{\chi t}, A^0_{\chi t}, A^0_{\chi \chi}} \simeq 126 \,\GeV$ if we allow for fine tuning at the level of ${\cal O}(126^2/(2m^2_{t\chi}/\epsilon^2)) \simeq10^{-5}$. 

\section{The constraints from the electroweak precision test and $Z \to b_L\bar{b}_L$}
\label{EWPTandZbb}

\begin{figure}[htbp]
\begin{center}
\begin{tabular}{cccc}
{
\begin{minipage}[t]{0.22\textwidth}
\includegraphics[scale=0.5]{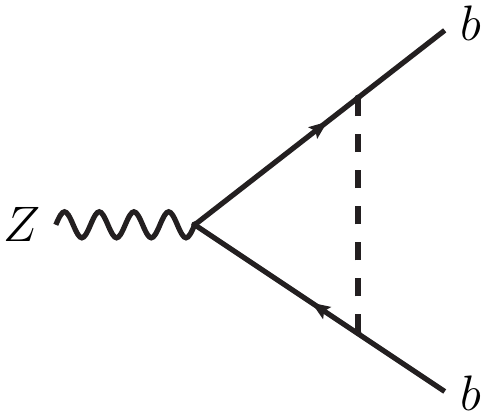} 
\end{minipage}
}
&
{
\begin{minipage}[t]{0.22\textwidth}
\includegraphics[scale=0.5]{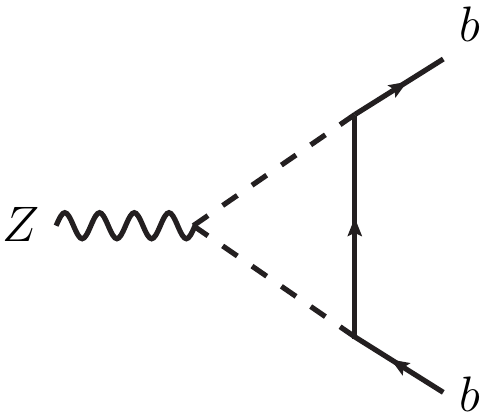} 
\end{minipage}
}
&
{
\begin{minipage}[t]{0.22\textwidth}
\includegraphics[scale=0.5]{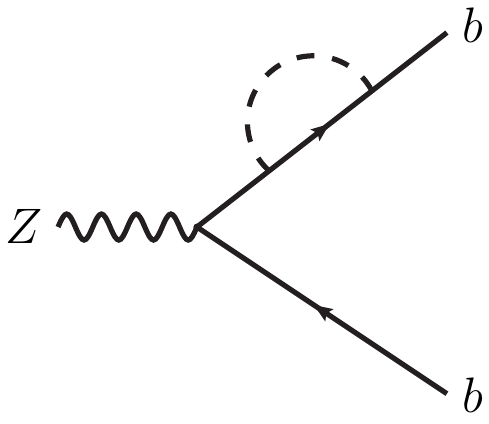} 
\end{minipage}
}
&
{
\begin{minipage}[t]{0.22\textwidth}
\includegraphics[scale=0.5]{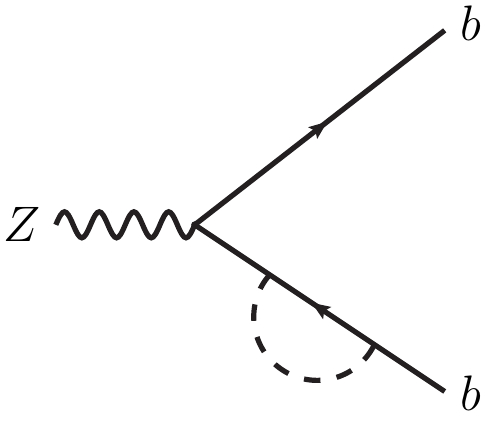} 
\end{minipage}
}
\end{tabular}
\caption[]{
Feynman diagrams for the would-be Nambu-Goldstone bosons contribution to $\delta g^b_L$. 
\label{Zbb-NGB}}
\end{center}
\end{figure}

\begin{figure}[htbp]
\begin{center}
\includegraphics[scale=1]{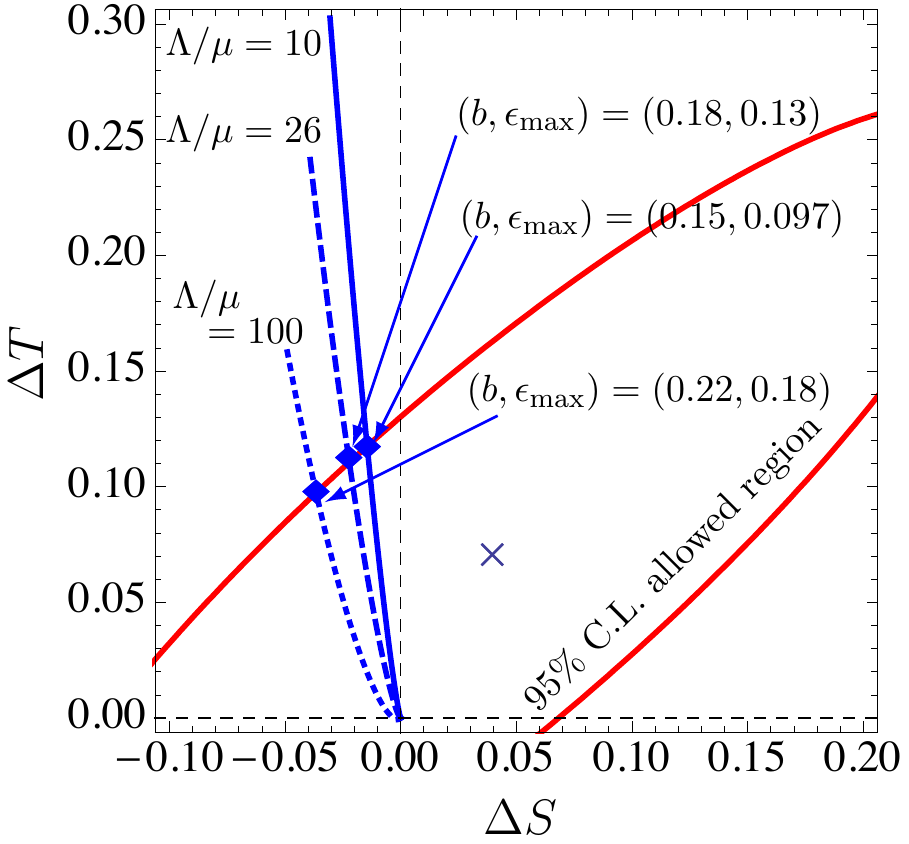} 
\caption[]{
$\Delta S, \Delta T$-constraints for the present top-seesaw model. The ellipsis (red) is the $95 \%\cl$ allowed region and $\times$ shows the central values given in Eq.(\ref{ST-central}) \cite{Beringer:1900zz}. The solid, dashed, dotted curves correspond to $\Lambda/\mu = 10$, 26 and 100, respectively. 
The symbols $\blackdiamond$ show the upper bound for $\epsilon$, i.e. $\epsilon < 0.097$ for $\Lambda/\mu = 10$, $\epsilon < 0.13$ for $\Lambda/\mu =26$ and $\epsilon < 0.22$ for $\Lambda/\mu = 100$. For reference, we also show the values of $b$ which are determined by the top quark mass for each case (see Eq.(\ref{typical-values})). We are assuming $a = 1$ here.
\label{TSS-ST}}
\end{center}
\end{figure}

In this section, we discuss the electroweak precision test constraints on our top-seesaw model. 
As we have discussed, four of the composite scalars ($H^\pm,H^\pm_{\chi b},H^0,A^0$) are very heavy 
and decouple from the low energy phenomenology. Furthermore, four neutral scalars ($H^0_{\chi t},H^0_{\chi \chi},A^0_{\chi t}, A^0_{\chi \chi}$) consist dominantly of the $SU(2)_L$ singlet scalar field. Thus, we include only $h^0$ in the analysis. However, the present $h^0$ can be identified with the SM Higgs boson and so its contribution is already included in the SM results i.e. we do not take into account the $h^0$ contribution when we consider the constraints on the Peskin-Takeuchi $S$ and $T$ parameters  \cite{Peskin:1990zt,Peskin:1991sw}, i.e. $\Delta S \equiv S -S_{\rm SM}$ and similarly for $\Delta T$. 
Therefore, we only take into account the contribution from the vector-like top quark partner ($\chi$) to $S$, $T$ and $R_b \equiv \Gamma(Z\to \bar{b}_Lb_L)/\Gamma(Z \to \text{hadrons})$ in our top-seesaw model. To compute these contributions, we should take into account the interactions between the quarks and the electroweak gauge bosons in the quark mass basis, Eqs.(\ref{rotate-L}) and (\ref{rotate-R}), and it is given by 
\beq
{\cal L}_{Vff}
\!\!&=&\!\!
\left[ 
\frac{1}{\sqrt{2}} g c^t_L W^+_\mu 
\left( c^t_L\bar{t}_L \gamma^\mu b_L + s^t_L\bar{\chi}_L \gamma^\mu b_L \right)
+
\text{h.c.}
\right]
\nonumber\\
&&
+
\frac{1}{2} (g W^3_\mu - g'B_\mu )
\left[ 
(c^t_L)^2 \bar{t}_L \gamma^\mu t_L + (s^t_L)^2 \bar{\chi}_L \gamma^\mu \chi_L 
+ (c^t_Ls^t_L) \left( \bar{t}_L \gamma^\mu\chi_L + \bar{\chi}_L \gamma^\mu t_L \right)
- \bar{b}_L \gamma^\mu b_L
\right]
\nonumber\\
&&
+
g' B_\mu 
\left[ 
Q_t  \bar{t} \gamma^\mu t + Q_t  \bar{\chi} \gamma^\mu \chi + Q_b  \bar{b} \gamma^\mu b
\right]
\,,\label{Vff-TSS}
\eeq 
where $W^{\pm,3}_\mu, B_\mu$ are the $SU(2)_L$ and $U(1)_Y$ gauge bosons, $g,g'$ are the $SU(2)_L$ and $U(1)_Y$ gauge couplings and $Q_t,Q_b$ are the $U(1)_{e.m.}$ charge of the top quark $Q_t = 2/3$ and the bottom quark $Q_b =-1/3$. %
Thus, the vector-like quark contribution to the $S,T$-parameters \cite{Lavoura:1992np,Maekawa:1995ha} in the top-seesaw model are given by
\beq
\Delta S 
&=&
 \frac{N_c}{2\pi} (s^t_L)^2
\left[ 
-\frac{1}{9} \ln \frac{m^2_\chi}{m^2_t} - (c^t_L)^2 f(m^2_\chi, m^2_t)
\right]
\,,\label{S-TSS}\\
\Delta T 
&=&
 \frac{N_c m^2_t}{16\pi s^2_W c^2_W M^2_W} (s^t_L)^2
\left[ 
(s^t_L)^2\frac{m^2_\chi}{m^2_t}
- \left(1+(c^t_L)^2\right) 
+ 2(c^t_L)^2 \frac{m^2_\chi}{m^2_\chi - m^2_t} \ln \frac{m^2_\chi}{m^2_t} 
\right]
\,,\label{T-TSS}
\eeq
where $f(m^2_1,m^2_2)$ are given by
\beq
f(m^2_1, m^2_2)
=
\frac{5(m^4_1 + m^4_2) - 22 m^2_1 m^2_2}{9(m^2_1 - m^2_2)^2}
+
\frac{3m^2_1 m^2_2 (m^2_1 + m^2_2) -m^6_1 - m^6_2}{3(m^2_1 - m^2_2)^3}
\ln \frac{m^2_1}{m^2_2}
\,.
\eeq
Next, let us consider the $Z\bar{b}_L\bar{b}_L$ constraint on the $\delta g^b_L$. Generally, the radiative correction of $\delta g^b_L$ is defined as
\beq
\frac{g}{c_W} Z_\mu \bar{b}_L [g^b_L + \delta g^b_L] b_L
\,,
\eeq
where $c_W \equiv \cos \theta_W$ is the cosine of the Weinberg angle and $g^{t,b}_{L,R}$ are given by
\beq
&&
g^t_L = \frac{1}{2} - \frac{2}{3} s^2_W
\quad , \quad
g^t_R = -\frac{2}{3} s^2_W
\,,\\
 &&
g^b_L = -\frac{1}{2} + \frac{1}{3} s^2_W
\quad , \quad
g^b_R = \frac{1}{3} s^2_W
\,.
\eeq
The difference between the present top-seesaw model and the SM arises from the Yukawa sector, and the largest one-loop contribution to $\delta g^b_L$ comes from the Yukawa interaction between the top quark, its vector-like partner and the would-be Nambu-Goldstone bosons. Hence it is sufficient to compute Feynman diagrams in Fig.\ref{Zbb-NGB}. %
In the SM, the contribution from Fig.\ref{Zbb-NGB} to $\delta g^b_L$ is given by
\beq
\left.\delta g^b_L\right|_{\rm SM} = \frac{m^2_t}{16\pi^2 v^2_{\rm EW}}
\,.\label{SM-gbL}
\eeq
In the present top-seesaw model the Yukawa term in Eq.(\ref{eff-Lag}) is given by
\beq
{\cal L}_{Gff}
&\supset&
y  \left( c^t_R \sin \alpha + s^t_R \cos \alpha \cos \theta\right) \cdot G^- \bar{b}_L t_R  
\nonumber\\
&&
+
y \left( -s^t_R \sin \alpha + c^t_R \cos \alpha \cos \theta\right) \cdot G^- \bar{b}_L \chi_R  
+
\text{h.c.}\,.
\eeq
Then the diagrams in Fig.\ref{Zbb-NGB} lead to the contribution to $\delta g^b_L$ which is given by
\beq
\left. \delta g^b_L \right|_{\rm TSS}
&=&
\frac{1}{16\pi^2} \frac{y^2/2}{(1+a^2b^2)(1+b^2)} \times
\nonumber\\
&&
\left[
\begin{aligned}
&
\cos^2\theta + a^2b^2
\\
&
- (s^t_L)^2 \left( c^t_R ab + s^t_R \cos\theta\right)^2
- (c^t_L)^2 \left( s^t_R ab - c^t_R \cos\theta\right)^2
\\
&
+
2 (c^t_L s^t_L) 
\left( c^t_R ab + s^t_R \cos\theta\right) \left( -s^t_R ab + c^t_R \cos\theta\right)
\frac{m_\chi m_t}{m^2_\chi - m^2_t} \ln \frac{m^2_\chi}{m^2_t}
\end{aligned}
\right]\,. \label{TSS-gbL}
\eeq
Then we rewrite Eqs.(\ref{S-TSS}), (\ref{T-TSS}) and (\ref{TSS-gbL}) in terms of parameters $a$, $b$ and $\epsilon$ (see Eq.(\ref{ratio-vev})). To the lowest order in $\epsilon^2$, the resultant expressions are 
\beq
\Delta S
&=&
\frac{N_c}{2\pi} \epsilon^2 \left[\frac{5}{9} + \frac{5}{9} \ln(b\epsilon) \right]
\,,\label{S-TSS-fin}\\
\Delta T
&=&
\frac{N_c}{16\pi^2s^2_Wc^2_W} \frac{m^2_t}{M^2_W} \frac{\epsilon^2}{b^2} \left[1 -2b^2 + 4b^2 \ln(b\epsilon) \right]
\,,\label{T-TSS-fin}\\
\left.\delta g^b_L \right|_{\rm TSS}
&=&
\frac{1}{16\pi^2}\frac{y^2}{2} b^2(a+\cos\theta)^2
\left[ 1-2\epsilon^2 \frac{\cos\theta}{a+\cos \theta} \ln(b\epsilon) \right]
\,.\label{TSS-gbL-fin}
\eeq
The experimental results for $\Delta S$ and  $\Delta T$ are \cite{Beringer:1900zz}
\beq
\Delta S = 0.04 \pm 0.09
\quad, \quad 
\Delta T = 0.07 \pm 0.08
\,,\label{ST-central}
\eeq
and the $95 \% \cl$ constraint on $\Delta g^b_L \equiv \delta \left.g^b_L\right|_{\rm TSS}-\left.g^b_L\right|_{\rm SM}$ is given by \cite{Dawson:2012di}
\beq
-2.7 \times 10^{-3} < \Delta g^b_L < 1.4 \times 10^{-3}
\,.\label{Zbb-95constraint}
\eeq
In Fig.\ref{TSS-ST}, we show $\Delta S,\Delta T$ in the present top-seesaw model for $\Lambda/\mu = 10$ (blue, solid curve), $26$ (blue, dashed curve) and $100$ (blue, dotted curve) together with the $95 \%\cl$ ellipsis (solid, red) on the $(\Delta S, \Delta T)$-plane. The diamond symbols on the curves give the upper bound for $\epsilon$. Thus we find that Fig.\ref{TSS-ST} shows $\epsilon$ should satisfy $\epsilon \lessim 0.1-0.2$. This result ensures that it was consistent to neglect ${\cal O}(\epsilon^2)$-terms in section \ref{TSS-review}. Furthermore, $\epsilon$ should satisfy Eq.(\ref{epsilon-lowerbound}), which implies that for $\Lambda/\mu = 10,26,100$, $\epsilon$ should satisfy $\epsilon > 0.34, 0.11, 0.024$, respectively. Thus, to maintain the self-consistency, $\Lambda/\mu = \Lambda/m_\chi$ should have a lower bound
\beq
\frac{\Lambda}{m_\chi}
\gtrsim 26\,.
\eeq
For $\epsilon \lessim 0.13$ and $\Lambda/\mu \gtrsim 26$, we obtain $\Delta g^b_L \simeq 4-7 \times 10^{-4}$ and this value satisfies the constraint from Eq.(\ref{Zbb-95constraint}) on $R_b$.
The values of the vector-like top quark partner mass ($m_\chi$) and the dynamical mass($m_{t\chi}$) in the present top-seesaw model with $b =0.18, \epsilon=0.13,\Lambda/\mu = 26$ are given as $m_\chi \simeq 3.8 \,\TeV$ and $m_{t\chi} \simeq 480\, \GeV$, respectively.

\section{Higgs bosons in the top-seesaw model at the LHC}
\label{VSLHC}

\subsection{The phenomenological Lagrangian for Higgs bosons  in the top-seesaw model}

To compare the Higgs bosons ($h^0\,, H^0_{\chi t}\,,A^0_{\chi t}\,,A^0_{\chi \chi}$) whose mass is smaller than $2M_W$ in the present top-seesaw model with the current LHC data, we first extract the relevant Higgs boson part from the effective Lagrangian Eq.(\ref{eff-LagV1}), and obtain
\beq
{\cal L}
&=&
c_{SVV} \left[ 
\frac{g^2 v_{\rm EW}}{2} W^{+\mu} W^-_{\mu} S
+
\frac{g^2 v_{\rm EW}}{4c_W} Z^\mu Z_\mu S
\right]
\nonumber\\
&&
- \sum^{\text{CP-even}}_{f=t,b,\tau,\chi}c_{Sff} \frac{m_f}{v_{\rm EW}} \bar{f}f S
- i \sum^{\text{CP-odd}}_{f=t,b,\tau,\chi}c_{Sff} \frac{m_f}{v_{\rm EW}} \bar{f}\gamma_5f S
\nonumber\\
&&
+
c_{Sgg} \frac{\alpha_s}{16\pi v_{\rm EW}} G^{\mu \nu}G_{\mu \nu} S
+
c_{S\gamma\gamma} \frac{\alpha}{8\pi v_{\rm EW}} F^{\mu \nu}F_{\mu \nu} S
\,.\label{Lag-LHC-higgs}
\eeq
Here $S=h^0,H^0_{\chi t},A^0_{\chi t},A^0_{\chi \chi}$, $\alpha_s \equiv g^2_s/(4\pi)$ and $\alpha \equiv e^2/(4\pi)$. 

The coefficients $c_{SVV}\,,c_{Sbb}$ and $c_{\tau\tau}$ are given by
\beq
c_{SVV}
&=&
\begin{cases}
\dfrac{v_{\rm TSS}}{v_{\rm EW}} 
\quad
&
\text{for $S=h^0$}
\\[1ex]
0
\quad
&
\text{for $S=H^0_{\chi t},A^0_{\chi t}, A^0_{\chi \chi}$}
\end{cases}
\,,\label{cSVV}
\\[1ex]
c_{Sbb}
&=&
\begin{cases}
\dfrac{m^{\rm TSS}_b}{m_b}
\quad
&
\text{for $S=h^0$}
\\[1ex]
0
\quad
&
\text{for $S=H^0_{\chi t},A^0_{\chi t},A^0_{\chi \chi}$}
\end{cases}
\,,\label{cSbb}
\\[1ex]
c_{S\tau\tau}
&=&
\begin{cases}
\dfrac{m^{\rm TSS}_\tau}{m_\tau}
\quad
&
\text{for $S=h^0$}
\\[1ex]
0
\quad
&
\text{for $S=H^0_{\chi t},A^0_{\chi t},A^0_{\chi \chi}$}
\end{cases}
\,.\label{cStautau}
\eeq

We expect that $c_{SVV}$, $c_{Sbb}$ and $c_{S\tau\tau}$ are equal to zero for 
$S=H^0_{\chi t}, A^0_{\chi t},A^0_{\chi \chi}$, since these scalars are dominantly composed of 
the $SU(2)_L$ singlet scalars (see Eqs.(\ref{mixing-CPevenHiggs}) and (\ref{mixing-CPoddHiggs})). 
Therefore it is reasonable to assume that $H^0_{\chi t}$, $A^0_{\chi t}$ and $A^0_{\chi \chi}$ do not couple to the electroweak gauge bosons and couple only to the top quark and its vector-like partner at tree level 
as one can see easily from Eq.(\ref{eff-LagV1}). The present top-seesaw model explains the observed mass hierarchy of the SM massive fields including the electroweak gauge bosons only by the condensation $\vev{\bar{\chi}_R q_L} \neq 0$. Hence, in this model we have $v_{\rm TSS} = v_{\rm EW}\,, m^{\rm TSS}_b = m_b\,,m^{\rm TSS}_\tau = m_\tau$, and this implies that the coefficients $c^{\rm SM}_{hVV} = c^{\rm SM}_{hbb} = c^{\rm SM}_{h\tau\tau} = 1$ correspond to the Higgs interactions of the SM. 

Of course it is possible that other sources contribute to the masses of SM matter fields  e.g. the extended (walking) technicolor \cite{Fukano:2012qx,Fukano:2012nx}. To gain insight on the parameter values of this type of models favored by the LHC data, we treat $c_{h^0VV}$, $c_{h^0bb}$ and $c_{h\tau\tau}$ as variable parameters in the following, and perform a global fit to the LHC data. Allowing for such freedom does not affect the dynamical aspects of the present top-seesaw model. This is so since $c_{SVV}$ and $c_{S\tau\tau}$ do not contribute to the dynamics in the present top-seesaw model, and $c_{Sbb}$ which does contribute to the dynamics is characterized by $\sin \theta$ which is very small as shown in Eq.(\ref{typical-values}). One should keep in mind however, that the electroweak precision test constraints in the section.\ref{EWPTandZbb} on parameters in the present top-seesaw model do change if $c_{h^0VV}$ is varied from 1.

The coefficients $c_{Stt},c_{S\chi\chi}$ are given by 
\beq
c_{Stt}
&=&
\begin{cases}
\dfrac{v_{\rm EW}}{m_t} \dfrac{y\left( c^t_Lc^t_R \,ab + c^t_L s^t_R \cos \theta \right)}{\sqrt{2(1+a^2b^2)}} 
\quad
&
\text{for $S=h^0$}
\\[1ex]
\dfrac{v_{\rm EW}}{m_t} \dfrac{y}{\sqrt{2}} s^t_L c^t_R
\quad
&
\text{for $S=H^0_{\chi t},A^0_{\chi t}$}
\\[1ex]
\dfrac{v_{\rm EW}}{m_t} \dfrac{y}{\sqrt{2}} [-s^t_L s^t_R]
\quad
&
\text{for $S=A^0_{\chi \chi}$}
\end{cases}
\,,\label{cStt}
\eeq
\beq
c_{S\chi\chi}
&=&
\begin{cases}
\dfrac{v_{\rm EW}}{m_\chi} \dfrac{y\left( -s^t_Ls^t_R \,ab + s^t_L c^t_R \cos \theta \right)}{\sqrt{2(1+a^2b^2)}} 
\quad
&
\text{for $S=h^0$}
\\[1ex]
\dfrac{v_{\rm EW}}{m_\chi} \dfrac{y}{\sqrt{2}} c^t_L s^t_R
\quad
&
\text{for $S=H^0_{\chi t},A^0_{\chi t}$}
\\[1ex]
\dfrac{v_{\rm EW}}{m_\chi} \dfrac{y}{\sqrt{2}} c^t_L c^t_R
\quad
&
\text{for $S=A^0_{\chi \chi}$}
\end{cases}
\,,\label{cSchichi}
\eeq
where $y,c^t_L,c^t_R$ are given in Eqs.(\ref{TSS-yukawa}), (\ref{def-cLt}) and (\ref{def-cRt}). Differently from the parameters $c_{SVV}$, $c_{Sbb}$ and $c_{S\tau\tau}$, one can not treat the coefficients $c_{Stt}$, $c_{S\chi\chi}$ as variable parameters since  they are related to the dynamics in the present top-seesaw model and they are strictly constrained from the top quark mass Eq.(\ref{top-constraint}), the electroweak precision test and validity of the present analysis as discussed in the section.\ref{EWPTandZbb}. %

The coefficients $c_{Sgg},c_{S\gamma\gamma}$ for the dimension-5 interaction terms in the last line in 
Eq.(\ref{Lag-LHC-higgs}) which come from the one-loop triangle diagrams are given by
\beq
c_{Sgg}
&=&
\sum_f c_{Sff} A_{1/2}(\tau_f)
\,,\label{cSgg}
\\
c_{S\gamma\gamma}
&=&
c_{SVV} A_1(\tau_W)
+
\sum_f N_c Q^2_f c_{Sff} A_{1/2}(\tau_f)
\,,\label{cSgg}
\eeq
where $N_c = 3(1)$ for quarks (leptons), $\tau_i \equiv 4m^2_i/m^2_h$ and the functions $A_1(x)$ and $A_{1/2}(x)$ are defined as
\beq
A_1(x)&=&2+3x+3x(2-x)f(x)
\,,\\
A_{1/2}(x)&=&
\begin{cases}
2x[1+(1-x)f(x)] & \text{for $S=h^0,H^0_{\chi t}$}
\\[1ex]
2x f(x) & \text{for $S=A^0_{\chi t},A^0_{\chi \chi}$}
\end{cases}
\,,\\
f(x) &=&
\begin{cases}
\left[ \arcsin (1/\sqrt{x})\right]^2 & \text{for $x>1$}
\\[1ex]
-\frac{1}{4} \left[ \ln \dfrac{1+\sqrt{1-x}}{1-\sqrt{1-x}} - i\pi\right]^2 & \text{for $x\leq 1$}
\end{cases}
\,.
\eeq

To compare with the current LHC data, it is appropriate to define the signal strengths in the present top-seesaw model as
\beq
\mu^{\rm TSS}_X(S)
= \left[ \sum_Y \zeta_Y \frac{\sigma_Y(S^0; {\rm TSS})}{\sigma_Y(h^0; {\rm SM})} \right]\times \frac{\left. \Br(S^0 \to X) \right|_{\rm TSS}}{\left. \Br(h^0 \to X) \right|_{\rm SM}}
\,,\label{def-signalstrength}
\eeq
where $\sigma_Y(S^0 ; \text{\rm TSS,SM})$ is the production cross section of the Higgs boson $S^0$$(= h^0$, $H^0_{\chi t}$, $A^0_{\chi t}$, $A^0_{\chi \chi})$ in the present top-seesaw model (TSS) and in the Standard Model (SM), $Y=$(GF, VBF, WH,ZH, ttH) refers to the production channel of the Higgs boson, $\left. \Br(S^0\to X) \right|_{\rm TSS,SM}$ is the branching ratio of the Higgs boson $S$ in TSS and SM, and $X$ denotes the decay channel of the Higgs bosons. Finally, $\zeta_Y$ is the efficiencies for the 
corresponding production channel and we assume that $\zeta_Y$ is the same for the TSS and SM cases. 
For the ratio of the production cross sections, we use the leading order estimates to obtain
\beq
\frac{\sigma_{\rm GF}(S^0; {\rm TSS})}{\sigma_{\rm GF}(h^0; {\rm SM})}
&=&
\left| \frac{c_{Sgg}({\rm TSS})}{c_{hgg}({\rm SM})} \right|^2
\,,\\[1ex]
\frac{\sigma_{\rm VBF}(S^0; {\rm TSS})}{\sigma_{\rm VBF}(h^0; {\rm SM})}
&=&
\frac{\sigma_{\rm WH}(S^0; {\rm TSS})}{\sigma_{\rm WH}(h^0; {\rm SM})}
=
\frac{\sigma_{\rm ZH}(S^0; {\rm TSS})}{\sigma_{\rm ZH}(h^0; {\rm SM})}
= c^2_{SVV}
\,,\\[1ex]
\frac{\sigma_{\rm ttH}(S^0; {\rm TSS})}{\sigma_{\rm ttH}(h^0; {\rm SM})}
&=&
c^2_{Stt}
\,.
\eeq
The current LHC data is shown in Tables \ref{current-ATLAS} and \ref{current-CMS}. 
In this paper, we use the efficiencies reported for $\gamma\gamma$ in 
\cite{ATLAS-CONF-2012-091,ATLAS-CONF-2013-012,CMS-PAS-HIG-13-001}, 
for $ZZ$ (CMS) in (\cite{CMS-PAS-HIG-13-002}, for $b\bar{b}$ (CMS) in \cite{CMS-PAS-HIG-12-044} and for $\tau^+\tau^-$(CMS) in \cite{CMS-PAS-HIG-13-004}. We use $\sigma^{\rm SM}_Y\,,(Y= \text{GF,VBF,WH,ZH,ttH})$ as presented in \cite{production-crosssection-at-8TeV} for all categories. 

\begin{table}[htbp]
\begin{center}
\centering
\begin{tabular}[t]{|c|c|c|c|c|}
\hline
\multicolumn{5}{ |c| }{ATLAS}   \\ [1 pt] \hline \hline
 & Category & $\hat{\mu} (7 \,\TeV)$ & $\hat{\mu} (8 \,\TeV)$ & Ref.   \\ [1 pt] 
\hline \hline
\multirow{14}{*}{$\gamma\gamma$}
&  Unconverted,Central, low $p_{Tt}$  & $0.54 \pm  1.43$ & $0.88 \pm 0.705$&  \multirow{14}{*}{\cite{ATLAS-CONF-2012-091},\cite{ATLAS-CONF-2013-012}} \\[1pt]
& Unconverted, Central, high $p_{Tt}$  & $0.21^{+1.73}_{-1.91}$ &  $0.95^{+1.07}_{-0.91}$&    \\[1pt]
& Unconverted, Rest, low $p_{Tt}$ & $2.52 \pm 1.66$ &  $2.51^{+0.90}_{-0.73}$&\\[1pt]
& Unconverted, Rest, high $p_{Tt}$  & $10.4^{+3.70}_{-3.66}$ &  $2.69^{+1.35}_{-1.14}$&\\[1pt]
& Converted, Central, low $p_{Tt}$ & $6.09^{+2.57}_{-2.63}$ &  $1.4^{+1.01}_{-0.94}$&\\[1pt]
& Converted, Central, high $p_{Tt}$ & $-4.37 \pm 1.77$ & $2.0^{+1.50}_{-1.24}$&\\[1pt]
& Converted, Rest, low $p_{Tt}$ & $2.75 \pm 1.99$  &  $2.2^{+1.15}_{-0.96}$&\\[1pt]
& Converted, Rest, high $p_{Tt}$ & $-1.64^{+2.93}_{-2.79}$ & $1.3 \pm 1.26$&\\[1pt]
& Converted, Transition  &  $0.34 \pm 3.59$ & $2.81^{+1.66}_{-1.58}$&\\[1pt]
& di-jets  & $2.72 \pm 1.86$ & - &\\[1pt]
& Loose high mass di-jet  & - & $2.77^{+1.75}_{-1.39}$&\\[1pt]
& Tight high mass di-jet  & - & $1.6 \pm 0.73$&\\[1pt]
& Low mass di-jet & - & $ 0.32^{+1.70}_{-1.44}$&\\[1pt]
&$E^{\rm miss}_T$ &-  &$3.0^{+2.71}_{-2.13}$&\\[1pt]
&one lepton & - & $2.7^{+1.94}_{-1.63}$&\\[1pt] 
\hline 
$ZZ$ & inclusive & \multicolumn{2}{|c|}{$1.5 \pm 0.4$} & \cite{ATLAS-CONF-2013-013}
\\[1pt]
\hline
$WW$ & inclusive & \multicolumn{2}{|c|}{$1.01 \pm 0.31$} & \cite{ATLAS-CONF-2013-030}
\\[1pt]
\hline
$b\bar{b}$ & VH & \multicolumn{2}{|c|}{$-0.4 \pm 1.06$} & \cite{ATLAS-CONF-2012-161}
\\[1pt]
\hline
$\tau^+\tau^-$ & inclusive & \multicolumn{2}{|c|}{$0.7 \pm 0.7$} & \cite{ATLAS-CONF-2012-160}
\\[1pt]
\hline
\end{tabular}
\caption{
The ATLAS Higgs data. The signal strength of diphoton channel are read off from Fig.14 in \cite{ATLAS-CONF-2012-091} for $7\,\TeV$ and Fig.13 in \cite{ATLAS-CONF-2013-012} for $8\,\TeV$. The signal strength of $ZZ,WW,b\bar{b},\tau^+\tau^-$ channels are $7 + 8 \,\TeV$ rtesults and presented in \cite{ATLAS-CONF-2013-013},\cite{ATLAS-CONF-2013-030},\cite{ATLAS-CONF-2012-161} and \cite{ATLAS-CONF-2012-160}.
\label{current-ATLAS}}
\end{center}
\end{table}

\begin{table}[htbp]
\begin{center}
\centering
\begin{tabular}[t]{|c|c|c|c|c|}
\hline
\multicolumn{5}{ |c| }{CMS}   \\ [1 pt] \hline \hline
 & Category & $\hat{\mu} (7 \,\TeV)$ & $\hat{\mu} (8 \,\TeV)$ & Ref.   \\ [1 pt] 
\hline \hline
\multirow{10}{*}{$\gamma\gamma$} 
&  Untagged-0 & $3.84^{+1.66}_{-2.03}$ & $2.15^{+0.74}_{-0.95}$&  \multirow{10}{*}{\cite{CMS-PAS-HIG-13-001}} \\[1pt]
&  Untagged-1 & $0.15^{+1.05}_{-0.92}$ & $0.06^{+0.68}_{-0.71}$& \\[1pt]
&  Untagged-2 & $0.03 \pm 1.23$ & $0.31^{+0.49}_{-0.46}$& \\[1pt]
&  Untagged-3 & $1.44^{+1.53}_{-1.69}$ & $-0.37^{+0.80}_{-0.86}$& \\[1pt]
&  di-jet & $4.21^{+1.75}_{-2.31}$ & - & \\[1pt]
&  di-jet, Tight & - & $0.28^{+0.56}_{-0.71}$ & \\[1pt]
&  di-jet, Loose & - & $0.83^{+0.98}_{-1.0}$ & \\[1pt]
&  muon & - & $0.4^{+1.35}_{-1.78}$ & \\[1pt]
&  electron & - & $-0.65^{+1.9}_{-2.7}$ & \\[1pt]
&$E^{\rm miss}_T$& - & $1.91^{+2.28}_{-2.61}$ & \\[1pt]
\hline
\multirow{2}{*}{$ZZ$} 
&  Untagged &  \multicolumn{2}{|c|}{$0.85^{+0.33}_{-0.27}$}&  \multirow{2}{*}{\cite{CMS-PAS-HIG-13-002}} \\[1pt]
&  di-jet&  \multicolumn{2}{|c|}{$1.22^{+0.85}_{-0.58}$}&   \\[1pt]
\hline
$WW$
&  inclusive &  \multicolumn{2}{|c|}{$0.76 \pm 0.21$}& \cite{CMS-PAS-HIG-13-003} \\[1pt]
\hline
\multirow{3}{*}{$b\bar{b}$} 
&  Z($l^+l^-$)H &  \multicolumn{2}{|c|}{$1.57^{+1.16}_{-1.10}$}&  \multirow{3}{*}{\cite{CMS-PAS-HIG-12-044}} \\[1pt]
&  Z($\nu\nu$)H&  \multicolumn{2}{|c|}{$1.8^{+1.08}_{-1.00}$}&   \\[1pt]
&  WH&  \multicolumn{2}{|c|}{$0.69^{+0.91}_{-0.86}$}&   \\[1pt]
\hline
\multirow{3}{*}{$\tau^+\tau^-$} 
&  1 jet &  \multicolumn{2}{|c|}{$0.75^{+0.50}_{-0.52}$}&  \multirow{3}{*}{\cite{CMS-PAS-HIG-13-004}} \\[1pt]
&  2 jet&  \multicolumn{2}{|c|}{$1.39^{+0.60}_{-0.56}$}&   \\[1pt]
&  VH&  \multicolumn{2}{|c|}{$0.77^{+1.49}_{-1.42}$}&   \\[1pt]
\hline
\end{tabular}
\caption{
The CMS Higgs data. The signal strength of diphoton channel are the mass-fit MVA data and we read off them from Fig.8(a) in \cite{CMS-PAS-HIG-13-001}. The signal strength of $ZZ$,$WW$,$b\bar{b}$,$\tau^+\tau^-$ channels are $7 + 8 \,\TeV$ rtesults. We read off them from Fig.6 in \cite{CMS-PAS-HIG-13-002}, Fig.7 in \cite{CMS-PAS-HIG-12-044} and Fig.8 in \cite{CMS-PAS-HIG-13-004}, respectively. As to the signal strength of $WW$ channel, we use the results reported in \cite{CMS-PAS-HIG-13-003}.
\label{current-CMS}}
\end{center}
\end{table}

\subsection{126 GeV Higgs boson in the top-seesaw model}

\begin{figure}[htbp]
\begin{center}
\begin{tabular}{cc}
{
\begin{minipage}[t]{0.4\textwidth}
\includegraphics[scale=0.6]{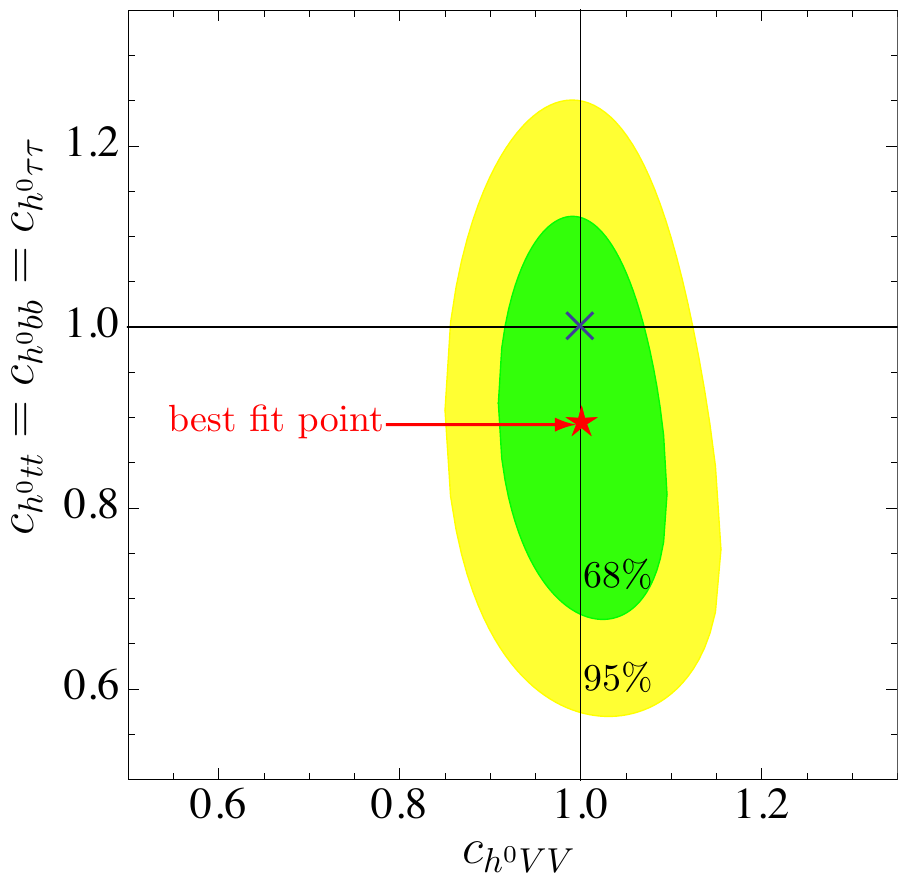} 
\end{minipage}
}
&
{
\begin{minipage}[t]{0.4\textwidth}
\includegraphics[scale=0.6]{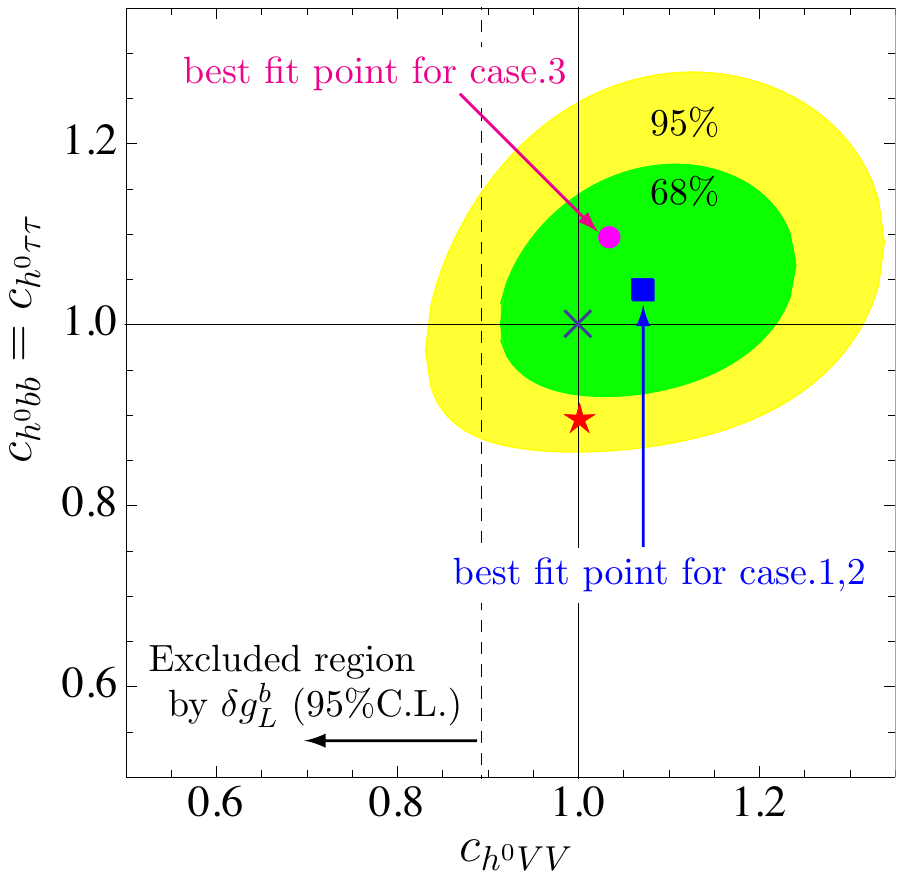} 
\end{minipage}
}
\end{tabular}
\caption[]{
Left  panel: $68 \%$ (green,inner), $95\%$ (yellow,outer) preferred region by the current LHC data on $(c_{h^0VV},c_{h^0ff})$-plane for $\chi^2$ of non-bias case. Right panel: $68 \%$ (green,inner), $95\%$ (yellow,outer) preferred region by the current LHC data on $(c_{h^0VV},c_{h^0ff})$-plane for $\chi^2$ of the top-seesaw model implementing case 1. The The vertical dashed line implies the lower bound by the $95 \% \cl$ constraint for the $\delta g^b_L$ and region on the left-hand side of this dashed line is excluded. The degenerate square symbols ({\scriptsize$\blacksquare$}, blue) and circle symbol ($\blackcircles$, magenta) correspond to Eqs.(\ref{TSScase1-min}),(\ref{TSScase2-min}) and Eq.(\ref{TSScase3-min}), respectively. In both panels, the star symbol ({\scriptsize$\bigstar$}, red), the cross ($\times$) are the best preferred values for non-bias case Eqs.(\ref{global-min}) and the SM value, respectively. 
\label{LHC-chisquare}}
\end{center}
\end{figure}

First, we consider preferred region on $(c_{h^0VV},c_{h^0ff})$-plane based on the current data presented in Tables \ref{current-ATLAS} and \ref{current-CMS}. 
We take $c_{h^0ff} \equiv c_{h^0bb} = c_{h^0\tau\tau} = c_{h^0tt}$ to compare with the SM Higgs boson case which corresponds to $c_{h^0VV} = c_{h^0_ff} =1$. 
For this purpose, we consider the $\chi^2$-function 
\beq
\chi^2 = \sum_i \left( \frac{\mu_{\rm th}-\mu_{\rm exp}}{\sigma_{\rm exp}}\right)^2
\,,\label{chisquare-fn}
\eeq
where $i$ corresponds to categories in Tables \ref{current-ATLAS} and \ref{current-CMS}, and 
$\mu_{\rm th}$ is the signal strength computed from the theory, $\mu_{\rm exp}$ is the observed signal strength and $\sigma_{\rm exp.}$ its $1 \sigma$ error. Without biasing model dependent constraints,  we find that the resultant $\chi^2$ minimum and the best fit value of $(c_{h^0VV},c_{h^0ff})$ are
\beq
\chi^2_{\min} = 58.87,  \quad c_{h^0VV}=1.00 \pm 0.09 , \quad c_{h^0ff} = 0.89^{+0.23}_{-0.22}
\,.\label{global-min}
\eeq
We show the no-bias preferred $68 \% \cl$ (inner, green) and $95 \% \cl$ (outer, yellow) regions on 
$(c_{h^0VV},c_{h^0ff})$-plane in the left panel of Fig.\ref{LHC-chisquare}. The star  symbol ({\scriptsize$\bigstar$}, red) and the cross symbol ($\times$) imply, respectively, the best fit values (Eq.(\ref{global-min})) and the  values corresponding to the SM Higgs boson which gives $\chi^2_{{\rm{SM}}} = 59.3$.

Next, let us consider the present top-seesaw model. 
In this paper, we do not consider a possibility that only the scalars arising dominantly from the 
$SU(2)_L$ singlet field may have $\simeq 126 \,\GeV$ mass. After all, such scalars do not couple to the electroweak gauge bosons at the tree level as shown in Eq.(\ref{cSVV}) and cannot provide for the observed $S \to WW,ZZ$ decay channels  \cite{ATLAS-CONF-2013-012,ATLAS-CONF-2013-030,CMS-PAS-HIG-13-002,CMS-PAS-HIG-13-003,ATLAS:2012ac,Aad:2012uub, Chatrchyan:2012dg,Chatrchyan:2012ty}. Therefore at least $h^0$ should be a candidate of the new boson with $\simeq 126 \,\GeV$. %

We  consider the following three potential scenarios for the new boson at $\simeq 126 \,\GeV$ in the top-seesaw model\footnote{see the discussion around Eqs.(\ref{mh0-g}), (\ref{mHchit0-g}), (\ref{mAchit0-g}) and (\ref{mAchichi0-g})}. 
\begin{description}
\item[case\,1 :] $m_{h^0} = 126 \, \GeV \ll m_{H^0_{\chi t}}, m_{A^0_{\chi t}}, m_{A^0_{\chi \chi}}$\,,
\item[case\,2 :] $m_{h^0} \simeq m_{A^0_{\chi \chi}} = 126 \, \GeV \ll m_{H^0_{\chi t}}, m_{A^0_{\chi t}}$\,,
\item[case\,3 :] $m_{h^0} \simeq m_{A^0_{\chi \chi}} \simeq m_{H^0_{\chi t}} \simeq m_{A^0_{\chi t}} = 126 \, \GeV$\,.
\end{description}
From now on, we will focus on each case in light of the current LHC data presented in Tables \ref{current-ATLAS} and \ref{current-CMS}. 
In comparison to the SM case discussed earlier, there are several noteworthy issues in the present top-seesaw model. First, we should  take $c_{h^0tt}$ independent of $c_{h^0ff} \equiv c_{h^0bb} = c_{h^0\tau\tau}$ since $c_{Stt}$ is defined by Eq.(\ref{cStt}) and it is  a function of $c_{h^0VV}$ through Eqs.(\ref{top-constraint}) and (\ref{cSVV}) once $\epsilon$ and $\Lambda/\mu$ are fixed. 
Also $m_\chi$ and $c_{S\chi\chi}$ are related to $c_{h^0VV}$ in analogous manner. 
Second, we should pay attention to the electroweak precision test constraints discussed in section \ref{EWPTandZbb} for $c_{h^0VV}$. Recall that we fixed $\Lambda/\mu =26\,,\epsilon=0.13$. However, we treat $b$ as a function of $c_{h^0VV}$ and require the value of $b$ to lead to realistic top quark mass, 
Eq. (\ref{top-constraint}), be compatible with the electroweak precision test at $95 \% \cl$ in 
Fig.\ref{TSS-ST} and be compatible with the $95 \% \cl$ constraint on $\delta g^b_L$ as given in 
Eq.(\ref{Zbb-95constraint}). One can see easily from Eq.(\ref{TSS-gbL-fin}) that $b$ becomes large for $v_{\rm TSS} < 246 \, \GeV$, i.e. $c_{h^0VV} < 1$. Thus, we find that $\delta g^b_L$ gives a lower bound for $c_{h^0VV}$ through the constraint on $b$ by Eq.(\ref{TSS-gbL-fin}). In fact, $\delta g^b_L$ constraint imposes $c_{h^0VV} \gtrsim 0.9$ for $\Lambda/\mu =26\,,\epsilon=0.13$. With these remarks in mind, let us now turn to the three cases listed above.

The case 1 is the simplest. In this case, out of the four Higgs bosons only $h^0$, originating from the $SU(2)_L$ doublet, is the lightest neutral scalar particle and its mass can be $\simeq 126 \,\GeV$ if $ g_{\chi t}  < 1 < g_\chi \simeq g_{\chi \chi}$ is satisfied in Eq. (\ref{mh0-g}). As we discussed below Eq. (\ref{mAchichi0-g}), this corresponds to fine tuning at the level of $10^{-3}$. In this case the signal strength of the Higgs boson is
\beq
\mu^{(1)}_X \equiv \mu^{\rm TSS}_X(h^0) \,.
\eeq
We find the minimum of $\chi^2$ 
to be 
\beq
\chi^2_{\min} \left[ \text{case 1} \right] = 58.9,\quad
c_{h^0VV}=1.07 ^{+0.17}_{-0.16}, \quad c_{h^0ff} = 1.04^{+0.14}_{-0.11}
\,,\label{TSScase1-min}
\eeq
which leads to the best fit values of $c_{h^0tt},c_{h^0\chi\chi},m_\chi$ as
\beq
c_{h^0tt} = 0.93 , \quad c_{h^0\chi\chi}=0.012, \quad m_\chi = 4119\, \GeV
\,.
\eeq

The case 2 can be realized by $g_{\chi t} < 1 \lessim g_\chi \simeq g_{\chi \chi}$ in Eqs. (\ref{mh0-g}) and (\ref{mAchichi0-g}). On the other hand, the case 3 is realized by $g_\chi \simeq g_{\chi \chi} \simeq g_{\chi t} \gtrsim 1$. Both cases are slightly different from the case 1 since $g_\chi \simeq 1$ is required,
and the level of fine tuning for both cases is estimated to be of the order of $10^{-5}$; see the discussion
below Eq. (\ref{mAchichi0-g}). Thus we obtain two light Higgs bosons with mass $m\simeq 126 \,\GeV$ for case 2 and four light Higgs bosons with 
mass $m\simeq 126 \,\GeV$ for case 3. %
In the case 3, we find that Eqs.(\ref{mHchit0-g}) and (\ref{mAchit0-g}) imply $m_{H^0_{\chi t}} \simeq m_{A^0_{\chi t}}$ since $\cos \theta \simeq 1$ and $b^2$ is very small. 
In the cases 2 and 3, we find that the $SU(2)_L$ singlet Higgs bosons $A^0_{\chi \chi}$,
$A^0_{\chi t}$ and $H^0_{\chi\chi}$ do not contribute to the process related to $W,Z,b,\tau$ 
as one can see easily from Eqs. (\ref{cSVV}), (\ref{cSbb}) and (\ref{cStautau}). However such 
Higgs bosons can be produced through the gluon fusion production process and they can decay to 
$\gamma \gamma$-channel thanks to non-zero couplings to the top quark and its vector-like partner quark as shown in Eqs.(\ref{cStt}) and (\ref{cSchichi}). Therefore, there are potentially extra contributions to the $\mu_{\gamma \gamma}$ via both the ratio of the production cross section and the branching ratio in Eq.(\ref{def-signalstrength}) in case 2 and case 3. Furthermore, several light Higgs bosons in the case 2 and the case 3 are nearly degenerate around at 126 GeV, and both cases realize the degenerate Higgs boson scenario considered in \cite{Gunion:2012gc,Batell:2012mj,Gunion:2012he,Ferreira:2012nv}. Hence the signal strengths are presented as
\beq
\mu^{(2)}_X \equiv  \mu^{\rm TSS}_X(h^0) +  \mu^{\rm TSS}_X(A^0_{\chi \chi})
\,,\label{2degenerate-mu}
\eeq
for case 2 and 
\beq
\mu^{(3)}_X \equiv  
\mu^{\rm TSS}_X(h^0) +  \mu^{\rm TSS}_X(A^0_{\chi \chi}) 
+ \mu^{\rm TSS}_X(H^0_{\chi t}) + \mu^{\rm TSS}_X(A^0_{\chi t}) 
\,,\label{4degenerate-mu}
\eeq
for case 3. Eqs. (\ref{2degenerate-mu}) and (\ref{4degenerate-mu}) are valid if the width of degenerate Higgs bosons are much smaller than the experimental mass resolution $\sigma_{\rm eff} \sim 1.5 \,\GeV$ \cite{ATLAS-CONF-2013-012,CMS-PAS-HIG-13-001}. In the present top-seesaw model, we find $\Gamma^{\rm TSS}_{\rm tot}(h^0) \simeq \Gamma^{\rm SM}_{\rm tot}(h^0) \simeq  10^{-3} \,\GeV$, $\Gamma^{\rm TSS}_{\rm tot}(A^0_{\chi \chi}) \simeq 10^{-6} \,\GeV$, $\Gamma^{\rm TSS}_{\rm tot}(A^0_{\chi t}) \simeq \Gamma^{\rm TSS}_{\rm tot}(H^0_{\chi t}) \simeq 10^{-4} \,\GeV$ and therefore Eqs.(\ref{2degenerate-mu}) and (\ref{4degenerate-mu}) are valid.

By using Eqs.(\ref{2degenerate-mu}) and (\ref{4degenerate-mu}), we find the minimum of $\chi^2$ function defined by Eq. (\ref{chisquare-fn}) as
\beq
\chi^2_{\min} \left[ \text{case 2} \right] = 58.9,\quad
c_{h^0VV}=1.07 ^{+0.17}_{-0.16}, \quad c_{h^0ff} = 1.04^{+0.14}_{-0.12}
\,,\label{TSScase2-min}
\eeq
for case 2 and
\beq
\chi^2_{\min} \left[ \text{case 3} \right] = 59.4,\quad
c_{h^0VV}=1.03 ^{+0.18}_{-0.16}, \quad c_{h^0ff} = 1.09^{+0.16}_{-0.13}
\,,\label{TSScase3-min}
\eeq
for case 3. At the best fit points, $c_{h^0tt},c_{h^0\chi\chi},m_\chi$ are 
\beq
c_{h^0tt} = 0.93 , \quad c_{h^0\chi\chi}=0.012, \quad m_\chi = 4114\,\GeV
\,,
\eeq
for case 2 and 
\beq
c_{h^0tt} = 0.96 , \quad c_{h^0\chi\chi}=0.012, \quad m_\chi = 3973\,\GeV
\,,
\eeq
for case 3.

We show the best fit values on the $(c_{h^0VV},c_{h^0ff})$-plane for the above three cases in  Fig.\ref{LHC-chisquare} right panel. The degenerate square symbols ({\scriptsize$\blacksquare$}, blue) and the circle symbol ($\blackcircles$, magenta) correspond to Eqs. (\ref{TSScase1-min}), (\ref{TSScase2-min}) and (\ref{TSScase3-min}), respectively. The shaded region  shows the preferred $68 \% \cl$ (inner, green) and $ 95 \% \cl$ (outer, yellow) regions for the case 1. For comparison, the best fit values for the no-bias result, Eq.(\ref{global-min}), and the SM result are also shown. The vertical dashed line shows the 
$\delta g^b_L$ $95 \% \cl$ lower bound for $c_{h^0VV}$ and the domain on the left-hand side of the dashed line is excluded at $95 \% \cl$ by the $\delta g^b_L$ constraint. 

The current LHC data in Tables \ref{current-ATLAS} and \ref{current-CMS} is compatible with the hypothesis that the new boson is the SM Higgs boson as shown in the left panel in Fig.\ref{LHC-chisquare}. 
From the right panel in Fig.\ref{LHC-chisquare} we observe that the $\delta g^b_L$ constraint imposes a strict constraint on $c_{h^0VV}$ in comparison to the constraint from the Higgs boson direct search. %
We find that the best fit points of the present top-seesaw model ($\chi^2_{\min}[\text{case 1}] \simeq \chi^2_{\min}[\text{case 2}]=58.9$) are slightly better than the SM ($\chi^2_{\min.}[\text{SM}] =59.3$).
However, when we impose further constraints leading to $c_{h^0VV} , c_{h^0bb}, c_{h^0\tau\tau} \leq 1$, we find that the best fit value for case 1 or case 2 is $c_{h^0VV} = c_{h^0bb} = 1$ and $c_{h^0\tau\tau} = 0.98$ at which we obtain $c_{h^0tt} = 0.99$ and $m_\chi = 3840 \,\GeV$. This underlines the fact that it will be important to see the precise measurement of the new boson couplings to fermions at the LHC and/or future linear collider experiments. At present, it is possible to identify  the new boson at the LHC with the potentially light Higgs boson(s) in the top-seesaw model. 

Possible alternative models incorporating similar strong dynamics and which could be compatible
with the current LHC data include topcolor
assisted technicolor \cite{Hill:1994hp} or the top-seesaw assisted technicolor \cite{Fukano:2012qx,Fukano:2012nx}.

\section{Summary}
\label{summary}

The present experimental situation at LHC can be interpreted in light of the naturality paradigm in  two 
different ways: First, it can be that the prevailing naturality paradigm is simply incompatible with Nature. 
The observable particle physics consists essentially of the Standard Model (SM), which is a consistent theory up to high energies implying that we live in a metastable universe. The dark matter, neutrino masses and other flavor phenomena arise either from physics from scales above $\sim 10^{10}$ GeV or hidden sectors superweakly coupled with the SM. Second, the traditional naturality paradigm is correct and new physics apart from the already observed Higgs boson awaits discovery in the terascale. 

In this paper we have adopted the latter point of view and considered a novel model for dynamical electroweak symmetry breaking and able to also explain the mass patterns of the third generation SM matter fields. The essential model building insight arising from the LHC data was the necessity to include the whole third generation of quarks and leptons into the dynamical framework.

We determined the low energy spectrum of fermions and bosons of the model, and confronted the model with the constraints from oblique electroweak parameters $S$ and $T$ and also from the $Z$-boson decay width to bottom quarks, $R_b$. We also constrained the model in order for the spectrum to be compatible with the LHC discovery of a 126 GeV scalar boson with the quantum numbers of the SM Higgs.

Finally, we performed a global fit to the LHC Higgs data. We showed that the model is compatible with the current observations at similar precision as the SM itself. In comparison with the SM, our top-seesaw model provides explanation for the mass patterns within the third generation. It also predicts an interesting spectrum of fermionic and bosonic states which could be discovered in the future. We have outlined three scenarios with distinct low energy spectrum. The scenario we called case 1 resembles most closely the SM, and has a single light scalar in the spectrum. The scenarios we called case 2 and case 3 differ from SM since they feature additional light scalars possibly accessible in the future LHC data. The model features also electroweak singlet scalars which manifest only through their coupling with the top quark. Hence, phenomenological analysis of the top quark associated production of the Higgs boson could provide further constraints for the model. Also, accumulating data for the direct search of the top quark partner $t^\prime$, whose mass is expected to be of the order of ${\cal O}($TeV$)$, in $t^\prime\rightarrow Wb$ will provide further constraints.

\section*{Acknowledgments}
We thank S. Chivukula and K. Yamawaki for insightful correspondence and H.-J. He, C.T. Hill and M. Tanabashi for careful reading of the manuscript.
H.S.F is supported in part by the JSPS Grant-in-Aid for Scientific Research (S) \#22224003.

\appendix
\section{Derivation of the effective Lagrangian}
\label{efflagrderivation}
We start with the NJL Lagrangian describing the new physics and its effects on the full third generation of SM matter which is  defined at the cut-off scale $\Lambda$ as defined in the main text
\beq
{\cal L}^{\rm TSS}_{\Lambda}
&=&
- \left[ \mu_{\chi \chi} \, \bar{\chi}_R \chi_L + \mu_{\chi t} \, \bar{t}_R \chi_L + \text{h.c.} \right]
\nonumber\\
&&
+
G_t \left( \bar{q}^\alpha_L \, t_R \right)\left( \bar{t}_R \, q^\alpha_L\right)
+
G_{q b} \left( \bar{q}^\alpha_L \, b_R \right)\left( \bar{b}_R \, q^\alpha_L\right)
+
G_{q \chi} \left( \bar{q}^\alpha_L \, \chi_R \right)\left( \bar{\chi}_R \, q^\alpha_L\right)
\nonumber \\
&&
+
G_{\chi \chi} \left( \bar{\chi}_L \, \chi_R \right)\left( \bar{\chi}_R \, \chi_L\right)
+
G_{\chi t} \left( \bar{\chi}_L \, t_R \right)\left( \bar{t}_R \, \chi_L\right)
+
G_{\chi b} \left( \bar{\chi}_L \, b_R \right)\left( \bar{b}_R \, \chi_L\right)
\nonumber\\
&&
+
G_2  \left[ 
\Bigl( \bar{q}^\alpha_L \, \chi_R \Bigr) (i\tau_2)^{\alpha \beta}\left( \bar{b}_R \, q^\beta_L \right)^c 
-
\Bigl( \bar{q}^\alpha_L \, b_R \Bigr)^c (i\tau_2)^{\alpha \beta}\left( \bar{\chi}_R \, q^\beta_L \right) 
\right]
\nonumber\\
&&
+
G_\tau  \left[ 
\Bigl( \bar{q}^\alpha_L \, \chi_R \Bigr) (i\tau_2)^{\alpha \beta}\left( \bar{\tau}_R \, l^\beta_L \right)^c 
-
\Bigl( \bar{l}^\alpha_L \, \tau_R \Bigr)^c (i\tau_2)^{\alpha \beta}\left( \bar{\chi}_R \, q^\beta_L \right) 
\right]
\,.\label{starting-NJL}
\eeq

Before implementing the large-$N_c$ fermion loop approximation \cite{Bardeen:1989ds}, it is convenient to diagonalize $G_{q \chi}$, $G_{q b}$ and $G_2$-terms in 
Eq.(\ref{starting-NJL}) \cite{Harada:1990wg}: 
\beq
{\cal L}^{\rm TSS}_{\Lambda} && 
\supset
\bpm \bar{q}^\alpha_L \, \chi_R && \left( \bar{q}^\beta_L \, b_R \right)^c \!\!\!(-i\tau_2)^{\beta \alpha}\epm
\bpm G_{q \chi} & G_2 \\[2ex] G_2 & G_{q b}\epm
\bpm \bar{\chi}_R \, q^\alpha_L \\[1ex] (i\tau_2)^{\alpha \beta}\left( \bar{b}_R \, q^\beta_L \right)^c  \epm
\nonumber\\
&&
=
\bpm \bar{q}^\alpha_L \, \chi_R && \left( \bar{q}^\beta_L \, b_R \right)^c \!\!\!(-i\tau_2)^{\beta \alpha}\epm 
\times
\nonumber\\
&& \hspace*{10ex}
\bpm \cos \theta & -\sin\theta \\[2ex] \sin\theta & \cos\theta\epm
\bpm G_\chi & 0 \\[2ex] 0 & G_b \epm
\bpm \cos \theta & \sin\theta \\[2ex] -\sin\theta & \cos\theta\epm
\bpm \bar{\chi}_R \, q^\alpha_L \\[1ex] (i\tau_2)^{\alpha \beta}\left( \bar{b}_R \, q^\beta_L \right)^c  \epm
\,,\label{diag-G2}
\eeq
where the eigenvalues $G_{\chi,b}\,,(G_\chi > G_b)$ are
\beq
G_{\chi , b} = \frac{1}{2} \left[ G_{q \chi} + G_{q b} \pm  \sqrt{(G_{q \chi} - G_{q b})^2 + 4G^2_2}\right]
\,,\label{def-ev-4f}
\eeq
and $\cos\theta\,,\,\sin\theta$ are given by $(0 \leq \theta \leq \pi/2)$
\beq
\cos^2\theta = \frac{1}{2} \left[ 1 + \frac{G_{q \chi} - G_{q b}}{\sqrt{(G_\chi - G_b)^2 + 4G^2_2}}\right]
\,,\,
\sin^2\theta = \frac{1}{2} \left[ 1 - \frac{G_{q \chi} - G_{q b}}{\sqrt{(G_\chi - G_b)^2 + 4G^2_2}}\right]
\,.\label{diag-matrix}
\eeq
To relate the bottom quark mass and the condensation $\vev{\bar{\chi}_R q_L} \neq 0$, we assume $G_b = 0$, i.e. we impose $G_{q \chi} G_{q b} - G^2_2 =0$ in Eq. (\ref{starting-NJL}). Hence the Lagrangian Eq. (\ref{starting-NJL}) becomes
\beq
{\cal L}^{\rm TSS}_{\Lambda}
&=&
- \left[ \mu_{\chi \chi} \, \bar{\chi}_R \chi_L + \mu_{\chi t} \, \bar{t}_R \chi_L + \text{h.c.} \right]
\nonumber\\
&&
+
G_t \left( \bar{\psi}^\alpha_L \, t_R \right)\left( \bar{t}_R \, \psi^\alpha_L\right)
\nonumber \\
&&
+
G_{\chi \chi} \left( \bar{\chi}_L \, \chi_R \right)\left( \bar{\chi}_R \, \chi_L\right)
+
G_{\chi t} \left( \bar{\chi}_L \, t_R \right)\left( \bar{t}_R \, \chi_L\right)
+
G_{\chi b} \left( \bar{\chi}_L \, b_R \right)\left( \bar{b}_R \, \chi_L\right)
\nonumber\\
&&
+
G_\chi  \left[ 
\cos \theta \Bigl( \bar{\psi}^\alpha_L \, \chi_R \Bigr) 
+ \sin \theta \Bigl( \bar{\psi}^\beta_L \, b_R \Bigr)^c (-i\tau_2)^{\beta \alpha}
\right]
\nonumber\\
&&
\hspace*{25ex}
\times
\left[
\cos \theta \left( \bar{\chi}_R \, \psi^\alpha_L \right) 
+ 
\sin \theta (i\tau_2)^{\alpha \beta}\left( \bar{b}_R \, \psi^\beta_L \right)^c 
\right]
\nonumber\\
&&
+
G_\tau  \left[ 
\Bigl( \bar{q}^\alpha_L \, \chi_R \Bigr) (i\tau_2)^{\alpha \beta}\left( \bar{\tau}_R \, l^\beta_L \right)^c 
-
\Bigl( \bar{l}^\alpha_L \, \tau_R \Bigr)^c (i\tau_2)^{\alpha \beta}\left( \bar{\chi}_R \, q^\beta_L \right) 
\right]
\,.\label{starting-new-NJL}
\eeq
We also impose on $G_t$, $G_{\chi}$ and $G_{\chi b}$ the following criticality conditions
\beq
0 < G_t < G_\crit < G_{\chi}
\quad ,\quad
0 < G_{\chi b} < G_\crit 
\quad , \quad
0 < G_{\tau} \ll G_\crit
\,,\label{criticality-GAB}
\eeq
where $G^{-1}_\crit \equiv N_c \Lambda^2 /(8 \pi^2)$ is the critical four fermion coupling. 

Let us now apply the large-$N_c$ fermion loop approximation to Eq. (\ref{starting-new-NJL}). As the first step, we introduce the auxiliary fields corresponding to the bound states 
arising from the condensates of fermion bilinears appearing in the four fermion interactions
in Eq. (\ref{starting-new-NJL}). We expect that bound states can be formed only if $G \sim G_\crit$ is satisfied for the corresponding channel, and therefore we do not consider the bound states originating from $G_\tau$ term. The auxiliary fields are introduced by adding to Eq. (\ref{starting-new-NJL}) the terms
\beq
{\cal L}_{\rm aux} 
&=&
- 
\frac{1}{G_t} \left| \Phi^\alpha_t - G_t \left(\bar{t}_R q^\alpha_L\right) \right|^2
- 
\sum_{f=t,b,\chi}
\frac{1}{G_{\chi f}} \left| \phi_{\chi f} - G_{\chi f} \left(\bar{f}_R \chi_L\right) \right|^2
\nonumber\\
&&
- \frac{1}{G_\chi}
\left| \Phi^\alpha_\chi - G_\chi \left[ 
\cos \theta \left( \bar{\chi}_R \, q^\beta_L \right) 
+ 
\sin \theta (i\tau_2)^{\alpha \beta}\left( \bar{b}_R \, q^\beta_L \right)^c 
\right] \right|^2
\,,\label{add-aux}
\eeq
where $\Phi_{t,\chi}$ is $SU(2)_L$ doublet complex scalar field and $\phi_{\chi f} (f=t,b,\chi)$ is 
$SU(2)_L$ singlet complex scalar field. Thus we can rewrite the Lagrangian in Eq. (\ref{starting-NJL}) as
\beq
{\cal L}_\Lambda
&=&
{\cal L}^{\rm TSS}_\Lambda + {\cal L}_{\rm aux}
\nonumber\\
&=&
- \left[ \mu_{\chi \chi} \, \bar{\chi}_R \chi_L + \mu_{\chi t} \, \bar{t}_R \chi_L + \text{h.c.} \right]
\nonumber\\
&&
+\left[
\bar{q}^\alpha_L \Phi^\alpha_t t_R
+
\cos \theta \, \bar{q}^\alpha_L \Phi^\alpha_\chi \chi_R
+
\sin \theta \, \bar{q}^\alpha_L \tilde{\Phi}^\alpha_\chi b_R
+
r_\tau \bar{l}^\alpha_L \tilde{\Phi}^\alpha_\chi \tau_R
+
\sum_{f=t,b,\chi} \bar{\chi}_L \phi_{\chi f} f_R
+
\text{h.c.}
\right]
\nonumber\\
&&
-
\left[
\frac{1}{G_t}|\Phi_t|^2
+
\frac{1}{G_\chi}|\Phi_\chi|^2
+
\sum_{f=t,b,\chi} \frac{1}{G_{\chi f}}|\phi_{\chi f}|^2
\right]
\,,\label{Lag-with-aux}
\eeq
where $r_\tau \equiv G_\tau/G_\chi$ and $\tilde{\Phi}^\alpha_\chi \equiv (-i\tau_2)^{\alpha \beta}\Phi^{*\beta}_\chi$. 
\begin{figure}[htbp]
\begin{center}
\begin{tabular}{ccc}
{
\begin{minipage}[t]{0.3\textwidth}
\begin{flushleft} (a) \end{flushleft} \vspace*{-2ex}
\includegraphics[scale=0.3]{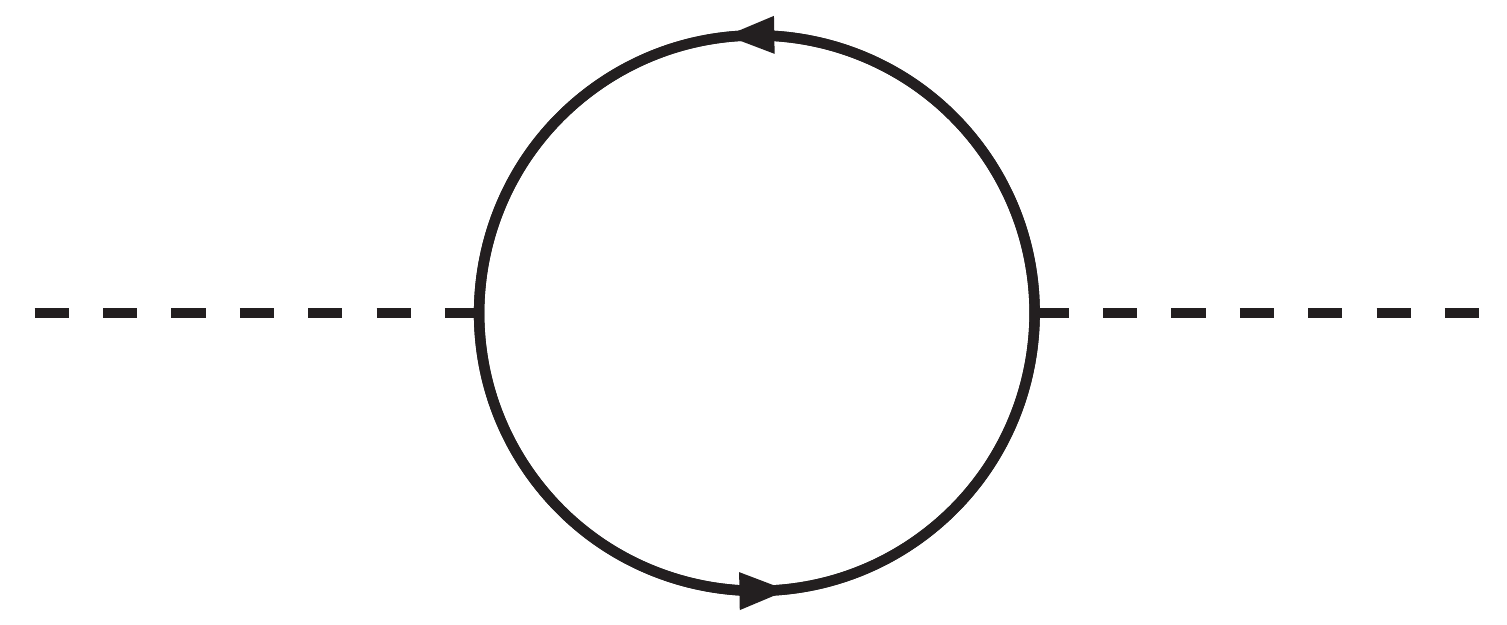} 
\end{minipage}
}
&
{
\begin{minipage}[t]{0.3\textwidth}
\begin{flushleft} (b) \end{flushleft} \vspace*{-2ex}
\includegraphics[scale=0.3]{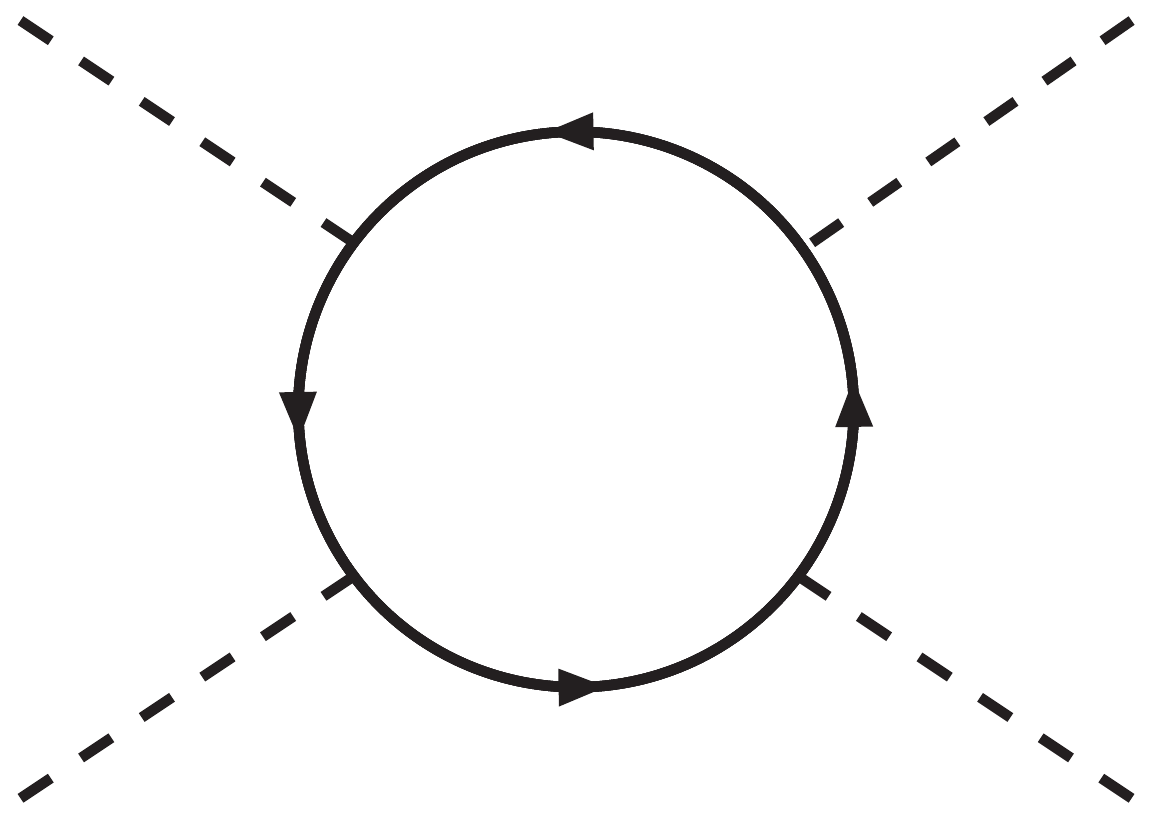} 
\end{minipage}
}
&
{
\begin{minipage}[t]{0.3\textwidth}
\begin{flushleft} (c) \end{flushleft} \vspace*{-2ex}
\includegraphics[scale=0.3]{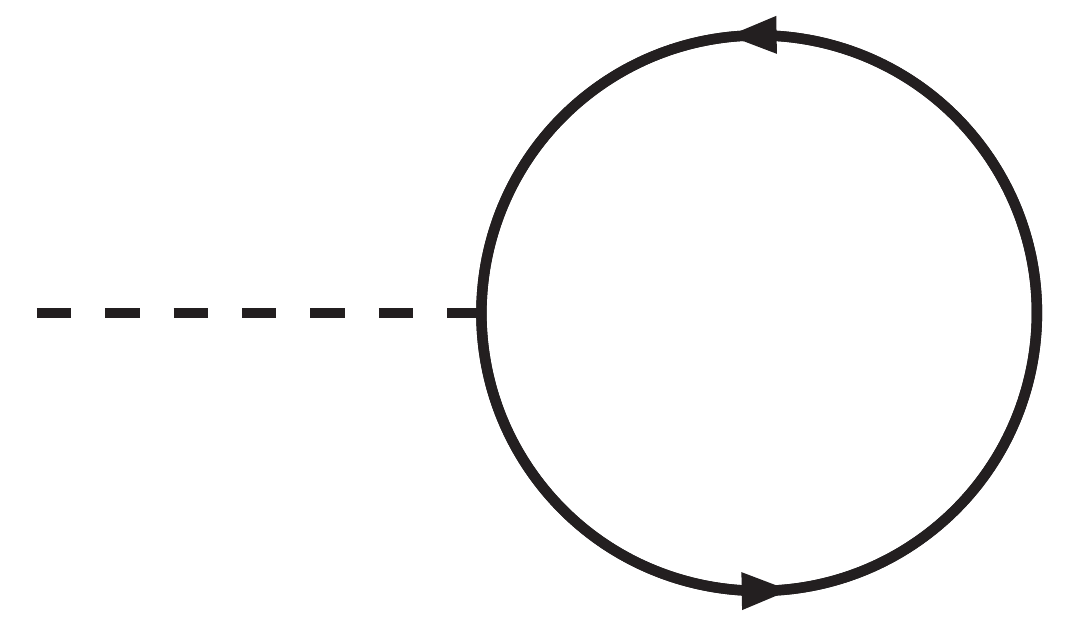} 
\end{minipage}
}
\end{tabular}
\caption[]{
Feynman diagrams for (a) kinetic term and mass term, (b) quartic coupling and (c) tadpole term for the dynamical scalar field $\Phi,\phi$ in the large-$N_c$ fermion loop approximation. The dashed and solid lines correspond to the fermion and scalar filed, respectively. 
\label{fermion-loop}}
\end{center}
\end{figure}%
As the second step, we compute the fermion loop as shown in Fig.\ref{fermion-loop}. 
The leptons do not contribute, and we obtain the effective Lagrangian 
\beq
{\cal L}^{\rm TSS}_{\mu < \Lambda}
&=&
- \left[ \mu_{\chi \chi} \, \bar{\chi}_R \chi_L + \mu_{\chi t} \, \bar{t}_R \chi_L + \text{h.c.} \right]
\nonumber\\
&&
+\left[
\bar{q}^\alpha_L \Phi^\alpha_t t_R
+
\cos \theta \, \bar{q}^\alpha_L \Phi^\alpha_\chi \chi_R
+
\sin \theta \, \bar{q}^\alpha_L \tilde{\Phi}^\alpha_\chi b_R
+
r_\tau \bar{l}^\alpha_L \tilde{\Phi}^\alpha_\chi \tau_R
+
\sum_{f=t,b,\chi} \bar{\chi}_L \phi_{\chi f} f_R
+
\text{h.c.}
\right]
\nonumber\\
&&
+
Z \left[ 
\left| D_\mu \Phi_t \right|^2 
+ \left| D_\mu \Phi_\chi \right|^2 
+ \sum_{f=t,b,\chi} \left| D_\mu \phi_{\chi f} \right|^2 
\right]
\nonumber\\
&&
- V(\Phi,\phi)
\,,\label{eff-Lag-0}
\eeq
where the covariant derivatives for $\Phi$ in the third line are the same as for the usual Higgs doublet field in the SM, while the covariant derivative for $\phi$ does not include the $SU(2)_L$ gauge interaction. The wave function renormalization of the scalar field, $Z$ in Eq. (\ref{eff-Lag-0}), comes from 
Fig.\ref{fermion-loop}(a) and is given by
\beq
Z = \frac{N_c}{16 \pi^2} \ln \frac{\Lambda^2}{\mu^2}
\,.\label{eq-Z}
\eeq
We will take $\mu = m_\chi$, where $m_\chi$ is the mass of the vector-like quark. Then our analysis is valid only for $\Lambda/\mu = \Lambda/m_\chi > m_\chi/m_t$. We will explicitly confirm that this requirement is satisfied. 

The effective potential, $V(\Phi,\phi)$ in Eq. (\ref{eff-Lag-0}), comes from Fig.~\ref{fermion-loop}~(a,b,c) and is given by
\beq
V(\Phi,\phi)
&=&
\left[ 
M^2_t \left| \Phi_t\right|^2 
+
M^2_\chi \left| \Phi_\chi\right|^2 
+
\sum_{f=t,b,\chi} M^2_{\chi f} \left| \phi_{\chi f}\right|^2
\right]
\nonumber\\
&&
+ C_{\chi t} \left[ \phi_{\chi t} + \phi^\dagger_{\chi t} \right]
+ C_{\chi \chi} \left[ \phi_{\chi \chi} + \phi^\dagger_{\chi \chi} \right]
\nonumber\\
&&
+\frac{\lambda}{2}
\left[ 
\begin{aligned}
&
\left( \Phi^\dagger_t \Phi_t + \phi^\dagger_{\chi t} \phi_{\chi t}  \right)^2
\\
&
+
\left( \sin^2\theta \Phi^\dagger_\chi \Phi_\chi + \phi^\dagger_{\chi b} \phi_{\chi b}  \right)^2
+
\left( \cos^2\theta \Phi^\dagger_\chi \Phi_\chi + \phi^\dagger_{\chi \chi} \phi_{\chi \chi}  \right)^2
\\
&
+
2 \left| \sin \theta \Phi^\dagger_t \tilde{\Phi}_\chi  + \phi^\dagger_{\chi t} \phi_{\chi b}\right|^2
+
2 \left| \cos \theta \Phi^\dagger_\chi \Phi_t  + \phi^\dagger_{\chi \chi} \phi_{\chi t}\right|^2
\\
&
+
2 \left| \sin\theta \cos \theta \tilde{\Phi}^\dagger_\chi \Phi_\chi + \phi^\dagger_{\chi b} \phi_{\chi \chi}\right|^2
\end{aligned}
\right]
\,,\label{eff-potential-0}
\eeq
where the parameters $M^2_A $ ($A=t,\chi,\chi t, \chi b, \chi \chi$), $C_B$ ($B=\chi t, \chi\chi$) and $\lambda$ are given by
\beq
M^2_A
&=&
\frac{N_c \Lambda^2}{8\pi^2} \left( \frac{1}{g_A} -1\right)
\,,\label{M2-0}
\\
C_B 
&=&
\frac{N_c \Lambda^2}{8\pi^2 g_B} \mu_B 
\,,\label{tadpole-0}
\\
\lambda
&=&
\frac{N_c}{8\pi^2} \ln \frac{\Lambda^2}{\mu^2} 
\,,\label{lambda-0}
\eeq
where we have defined $g_A$ as
\beq
g_A \equiv \frac{N_c \Lambda^2}{8 \pi^2} G_A
\,.\label{dimless-4f-def}
\eeq
Note that both $C_B$ and $\lambda$ are positive definite.
Let us then reconsider the criticality condition given in Eq.(\ref{criticality-GAB}). 
From the definition in Eq.(\ref{dimless-4f-def}), $G > G_\crit$ corresponds to $g > 1$ 
and $G < G_\crit$ to  $g < 1$. Thus, in the language of the effective potential, the supercritical coupling $G_A$ corresponds to $M_A^2<0$, while subcritical coupling $G_A$ corresponds to $M_A^2>0$. Therefore the criticality assumptions given in Eq.(\ref{criticality-GAB}) are equivalent to
\beq
M^2_\chi <0
\quad , \quad
M^2_t ,  M^2_{\chi b} > 0
\,,\label{criticality-eff-1}
\eeq
and $M^2_{\chi t, \chi\chi}$ are not constrained by the criticality conditions since we treat 
$G_{\chi t, \chi\chi}$ as free parameters. 
To normalize the kinetic term for the scalar field in Eq.(\ref{eff-Lag-0}), we rescale the fields $\Phi$ and 
$\phi$ as
\beq
\sqrt{Z} \Phi \to \Phi \quad,\quad \sqrt{Z} \phi \to \phi
\,,
\eeq
and eliminate $\mu_{\chi t,\chi \chi}$-terms by redefinition of $\phi_{\chi t, \chi \chi}$. Hence, the final form of the effective Lagrangian, valid for $\mu < \Lambda$, becomes
\beq
{\cal L}_{\mu < \Lambda}
&=&
\left[\left| D_\mu \Phi_t \right|^2 + \left| D_\mu \Phi_\chi \right|^2 + \sum_{f=t,b,\chi} \left| D_\mu \phi_{\chi f} \right|^2 \right]
\nonumber\\
&&
+y \left[
\bar{\psi}^\alpha_L \Phi^\alpha_t t_R
+
\cos \theta \, \bar{\psi}^\alpha_L \Phi^\alpha_\chi \chi_R
+
\sin \theta \, \bar{\psi}^\alpha_L \tilde{\Phi}^\alpha_\chi b_R
+
r_\tau \bar{l}^\alpha_L \tilde{\Phi}^\alpha_\chi \tau_R
+
\sum_{f=t,b,\chi} \bar{\chi}_L \phi_{\chi f} f_R
+
\text{h.c.}
\right]
\nonumber\\
&&
- V(\Phi,\phi)
\,,\label{eff-Lag}
\eeq
where $y$ is given by
\beq
y = \frac{1}{\sqrt{Z}}
=\frac{4\pi}{\sqrt{N_c \ln(\Lambda^2/\mu^2)}}
\,,\label{TSS-yukawaV1}
\eeq
and $V(\Phi,\phi)$ is given by Eq.(\ref{eff-potential-0}) with the replacements
\beq
&&
M^2_A \to 
\frac{M^2_A}{Z} = \frac{2\Lambda^2}{\ln(\Lambda^2/\mu^2)} \left( \frac{1}{g_A}-1\right)
\,,\label{M2-fin}\\
&&
C_B \to 
\frac{C_B}{\sqrt{Z}} = 
\mu_B \frac{\Lambda^2}{2\pi g_B} \sqrt{\frac{N_c}{\ln(\Lambda^2/\mu^2)}} 
\,,\label{tadpole-fin}\\
&&
\lambda \to 
\frac{\lambda}{Z^2} = \frac{32 \pi^2}{N_c \ln(\Lambda^2/\mu^2)} = 2y^2 
\,.\label{lambda-fin}
\eeq

\bibliography{126HiggsTSS.bib}

\end{document}